\documentclass[epj]{svjour}
\usepackage[numbers,sort&compress]{natbib}

\usepackage[dvipsnames]{xcolor}

\usepackage[T1]{fontenc}
\usepackage{ae,aecompl}

\usepackage{siunitx}
\DeclareSIUnit[]\sunmass{\text{\ensuremath{M_{\odot}}}}
\DeclareSIUnit[]\dyne{dyn}
\usepackage{bbm}

\usepackage{graphicx}   
\usepackage{amsmath}    
\usepackage{amssymb}    

\usepackage{fix-cm}

\usepackage{microtype}

\renewcommand{\d}[1]{\ensuremath{\operatorname{d}\!{#1}}}

\begin{document}

\title{Maximum mass and universal relations of rotating relativistic
  hybrid hadron-quark stars}

\titlerunning{Rotating relativistic hybrid stars}
\author{Gabriele
  Bozzola\inst{1}\thanks{Email: gabrielebozzola@email.arizona.edu}, Pedro L.\
  Espino\inst{2}\thanks{Email: pespino@email.arizona.edu}, Collin D. Lewin\inst{1}\thanks{Email: collinlewin@email.arizona.edu}, Vasileios
  Paschalidis\inst{1,2}\thanks{Email: vpaschal@email.arizona.edu}}
\institute{Department of Astronomy, University of Arizona, Tucson, Arizona, United States   \and
  Department of Physics, University of Arizona, Tucson, Arizona, United States
  }

\date{\today}

\abstract{We construct equilibrium models of uniformly and differentially
  rotating hybrid hadron-quark stars using equations of state (EOSs) with a
  first-order phase transition that gives rise to a third family of compact
  objects. We find that the ratio of the maximum possible mass of uniformly
  rotating configurations -- the supramassive limit -- to the
  Tolman-Oppenheimer-Volkoff (TOV) limit mass is not EOS-independent, and is
  between $1.15$ and $1.31$, in contrast with the value of $1.20$ previously
  found for hadronic EOSs. Therefore, some of the constraints placed on the EOS
  from the observation of the gravitational wave event GW170817 do not apply to
  hadron-quark EOSs. However, the supramassive limit mass for the family of EOSs
  we treat is consistent with limits set by GW170817, strengthening the
  possibility of interpreting GW170817 with a hybrid hadron-quark EOSs. We also
  find that along constant angular momentum sequences of uniformly rotating
  stars, the third family maximum and minimum mass models satisfy approximate
  EOS-independent relations, and the supramassive limit of the third family is
  approximately $\SI{16.5}{\percent}$ larger than the third family TOV limit. For
  differentially rotating spheroidal stars, we find that a lower-limit on the
  maximum supportable rest mass is $\SI{123}{\percent}$ more than the TOV limit rest
  mass. Finally, we verify that the recently discovered universal relations
  relating angular momentum, rest mass and gravitational mass for turning-point
  models hold for hybrid hadron-quark EOSs when uniform rotation is considered,
  but have a clear dependence on the degree of differential rotation.
}

\maketitle

\section{Introduction}
\label{sec:introduction}

The properties of the equation of state (EOS) of dense nuclear matter constitute
an important open problem that brings together several different fields ranging
from astrophysics and gravitation to particle and nuclear physics. While the EOS
is understood up to nuclear saturation density
$\rho_0 \approx \SI{2.7e15}{\g\per\cm\cubed}$ \cite{Haensel1994, Haensel2007,
  Lattimer2012}, the situation is less clear at super-nuclear densities, with
several different models being possible (see, e.g.~\cite{Heiselberg2000,
  Lattimer2012}). Regimes of such high density are expected to occur in Nature
in the central regions of compact stars, where the density can reach up to
several times $\rho_0$.

The EOS determines several of the macroscopic properties of compact stars, such
as mass and radius, thus affecting observable quantities (for reviews,
see~\cite{Lattimer2001, Hebeler2013, Lattimer2016, Ozel2016}). For this reason,
astronomical observation is one of the primary tools to probe the high-density
end of the EOS (see, e.g.,~\cite{Lattimer2007}). Recently, the detection of
gravitational waves from a binary neutron star~\cite{Abbott2017a} (event
GW170817), accompanied by electromagnetic counterparts (events GW170817-GBM, and
AT2017gfo~\cite{Abbott2017a, Abbott2017b, Abbott2017c,
  Fermi-LAT-Collaboration2017, Valenti2017, Savchenko2017}) provided a wealth of
information that, under certain assumptions, helped to place constraints on the
EOS~\cite{Bauswein2017b, Shibata2017, Margalit2017, Annala2018, Ruiz2018,
  Radice2018b, Abbott2018b, Raithel2018, Most2018, Rezzolla2018, Burgio2018,
  Koppel2019} (see also~\cite{Raithel:2019uzi} for a review). Among the EOSs
that are compatible with GW170817, there are those with a first-order phase
transition from hadronic to quark matter at densities larger than $\rho_0$, which
can lead to the so-called third family of stable stars. The third family
consists of objects that are more compact than white dwarfs and neutron stars.
These stars are hybrid hadron-quark stars (or just hybrid stars), i.e.,
configurations with a quark core surrounded by a hadronic shell.\footnote{It is
  also possible for a fourth family to arise \cite{Alford2017} if the EOS has
  an additional phase transition. However, this scenario will not be considered
  here.} The third family has been studied for a long time~\cite{Gerlach1968,
  Kampfer1981, Schertler1997, Schertler1998, Schertler2000, Glendenning2000,
  Ayvazyan2013,Blaschke2013,Benic2015,Haensel2016,Bejger2017,
  Kaltenborn2017,Alvarez-Castillo2017,Abgaryan2018,Maslov2018,Blaschke2018,Alvarez-Castillo2019},
with several of these earlier works focusing on ``twin stars'', which are
hybrid stars in the third family whose mass is in the same range as the mass of
neutron stars, but have smaller radii. Interest in hybrid stars has recently
increased due to GW170817, because GW170817 is also compatible with at least one
component being a hybrid star as first pointed out in~\cite{Paschalidis2018}.
Subsequent studies also focused on prospects of future multi-messenger
detections of the merger of binary compact stars with hybrid hadron-quark EOSs
(e.g.,~\cite{Drago2018,Burgio2018, Han2018, Chen2018, Annala2019, Most2019,
  Bauswein2019,DePietri:2019khb}).

Motivated by the fact that such hybrid stars are compatible with GW170817, in
this paper we focus on improving our understanding of uniformly and
differentially rotating hybrid stars. These have received only little
consideration \cite{Ayvazyan2013,Haensel2016} and many of their properties are
largely unexplored. We adopt the same EOSs with first-order phase transitions to
quark matter introduced in \cite{Paschalidis2018}, and focus on answering the
following set of questions: What is the maximum mass for uniformly and
differentially rotating hybrid stars? Are these compatible with constraints from
GW170817? Do rotating hybrid stars satisfy the same existing universal relations
that hadronic EOSs satisfy?

The maximum mass that a non-rotating relativistic star\footnote{We will
  interchangeably also use the terms ``static'' and ``TOV'' to indicate
  non-rotating stars \cite{Tolman1934,Oppenheimer1939}.} can support -- the TOV
limit -- is determined by the EOS. However, rotation can increase the maximum
supportable mass
\cite{Friedman1987,Cook1994,Eriguchi1994,PaschalidisStergioulas2017}. In the
case of hadronic EOSs, uniform rotation increases the maximum mass by about
$\SI{20}{\percent}$ \cite{Cook1994, Cook1994b, Lasota1996,
  Morrison2004,Breu2016}.\footnote{However, it has been shown that strange
  quark-matter stars can support up to $\SI{40}{\percent}$ more than the non-rotating
  limit \cite{Gourgoulhon1999,Rosinska2000,Bhattacharyya2016}.} Equilibrium
configurations whose mass exceeds the TOV limit are commonly referred to as
``supramassive'' \cite{Cook1992}. The maximum mass supported when allowing for
maximal uniform rotation is known as the ``supramassive limit''. Given the
approximate EOS-independence (for hadronic EOSs), the supramassive limit has
been used to place constraints on the TOV limit mass using observations
(e.g.~\cite{Most2018,Ruiz2018}). Whether this universality for the ratio of the
supramassive limit to the TOV limit mass holds for hybrid stars has not been
investigated before and we test this here. Using the code of~\cite{Cook1992}, we
construct mass shedding sequences and find that for the hybrid EOSs we treat,
this ratio is between $\num{1.15}$ and $\num{1.30}$. As a result, the
universality does not hold for hybrid stars or, at least, the spread of the
universality is significantly enhanced. However, using the same arguments as
in~\cite{Most2018,Ruiz2018}, we find that the supramassive limit of our hybrid
EOSs is consistent with GW170817, lending further support to the results
of~\cite{Paschalidis2018} that GW170817 is compatible with hybrid hadron-quark
EOSs.

Differential rotation can greatly boost the maximum mass that a star can have
\cite{Baumgarte2000,Morrison2004}. The term
``hypermassive''~\cite{Baumgarte2000} is used to describe those stars that are
more massive than the supramassive limit, and are supported by the additional
centrifugal support provided by differential rotation. There also exist stars
that can support more than two times the TOV limit mass (these were termed
``ubermassive'' in~\cite{Espino2019}), and which can arise only when
differential rotation is present. Differential rotation plays an important role
in temporarily stabilizing a binary neutron star merger remnant, and likely many
merger remnants go trough a hypermassive phase (see,
e.g.,~\cite{Baiotti2017,PaschalidisStergioulas2017,Paschalidis2017} for recent
reviews on binary neutron star mergers). Differentially rotating stars have been
shown to exhibit a rich and interesting solution space (see \cite{Ansorg2009,
  Studzinska2016, Rosinska2016} for polytropes, \cite{Espino2019} for hadronic
EOSs, and \cite{Zhou:2019hyy, Szkudlared2019} for strange quark stars). For
instance, as the degree of differential rotation varies, the topology of the
star can change from spheroidal to quasi-toroidal, where the maximum energy
density does not occur at the geometric center of the star, but in a ring around
the stellar center of mass. Studies of the dynamical stability of such
quasi-toroidal configuration are under way and it is found that these
configurations are dynamically unstable \cite{Espino2019b}. For this reason, in
this paper we focus primarily on the maximum mass that differentially rotating
\emph{spheroidal} hybrid star configurations can support using the differential
rotation law of~\cite{Komatsu1989a, Komatsu1989b}. The motivation for this study
is that quark deconfinement can happen not only for isolated stars, but also
following a binary neutron star merger \cite{Most2019}, that can result in a
differentially rotating hybrid star. Using the code of~\cite{Cook1992}, we find
that a lower-limit on the maximum supportable rest mass is $\SI{123}{\percent}$
more than the TOV limit rest mass. This means that ubermassive spheroidal hybrid
hadron-quark configurations can exist.

In the context of binary mergers, EOS-independent relations have proven useful
as tools to devise tests for general relativity and break degeneracies in
gravitational wave signals. For instance, the universal I-Love-Q relations
\cite{Yagi2013, Yagi2013b, Yagi2014, Yagi2017} were used in \cite{Abbott2018b}
to tighten the constraints placed on the tidal deformability of the two
components of the GW170817 binary. In \cite{Paschalidis2018} it was verified
that, for the EOSs we adopt here, the I-Love-Q relations are still satisfied
with a spread of at most $\SI{3}{\percent}$ in both slowly and rapidly rotating
hybrid stars.\footnote{Note that this result is in tension with what was found
  in \cite{Bandyopadhyay2018} where different equations of state were adopted.}
In \cite{Bozzola2018}, relations that do not depend on the EOS or the degree of
differential rotation were found for turning-point models, which are the
critical points of the gravitational mass along equilibrium sequences with fixed
angular momentum or rest mass. For uniformly rotating configurations, turning
points locate the onset of the unstable branch, and for this reason are useful
to study the stability properties of rotating stars \cite{Friedman1988,
  Takami2012, Bauswein2017, Weih2018}. We verify that the universal relations
relating angular momentum, rest mass and gravitational mass for turning-point
models found in \cite{Bozzola2018} hold for hybrid hadron-quark EOSs when
uniform rotation is considered, but have a clear, albeit weak, dependence on the
degree of differential rotation.

The rest of the paper is structured as follows. In
Section~\ref{sec:equations-state} we review the EOSs we consider in this work.
In Section~\ref{sec:methods} we briefly describe the code we use to perform our
study. In Section~\ref{sec:maxim-mass-supr} we compute the supramassive limit
for hybrid stars and discuss its implications for GW170817. We also test the
previously found universal relations of~\cite{Breu2016} and~\cite{Bozzola2018}
for the maximum mass along constant angular momentum sequences, and point out
that the minimum mass twin stars also satisfies a universal relation. In
Section~\ref{sec:maxim-mass-class} we compute the maximum rest mass of
differentially rotating spheroidal hybrid stars, and test the dependence on the
degree of differential rotation of the aforementioned universal relations. We
conclude in Section~\ref{sec:conclusions} with a summary of our main findings.
Unless otherwise specified, throughout this paper we adopt geometrized units
where $c = G = 1$ (where $c$ is the speed of light in vacuum and $G$ the
gravitational constant). Square brackets are used to designate the units of
quantities. For instance, $[\si{\sunmass}]$ indicates the solar mass. In
addition, when we use the word ``mass'' we refer to the gravitational
(Arnowitt-Deser-Misner -- ADM) mass for which we use the symbol $M$ (or
$M_{\rm ADM}$) with other sub- or superscripts except for the subscript $0$. We
will specify ``rest'' or ``baryon'' mass otherwise, and to designate rest mass
we use the subscript $0$ with other sub- or superscripts, e.g., $M_0$.

\section{Equations of State}
\label{sec:equations-state}

The hybrid hadron-quark EOSs we employ are the same as those developed in
\cite{Paschalidis2018} for the EOSs introduced in \cite{Alford2017} (Set I) and
\cite{Alvarez-Castillo2019} (Set II). Here we briefly describe these EOSs, and
refer the interested reader to \cite{Paschalidis2018} for more details. Note
that we use a different naming convention compared to \cite{Paschalidis2018},
and the map between the two is listed in Table~\ref{tab:eos}, where we also
report the maximum mass for static stars in both the hadronic and third family
branches.

The EOSs correspond to zero-temperature matter in beta-equilibrium with a
low-density hadronic phase and a high-density quark phase that are matched
through a first-order phase transition. High tension at the quark-hadron
interface is assumed, leading to a sharp transition boundary between the two
phases. A low-density crust component based on the model of \cite{Baym1971,
  Negele1973} is also added to both sets of EOSs. In all EOS models the pressure
matching between the two phases is performed via a Maxwell construction.

The Set I EOSs have a hadronic part that follows \cite{Colucci2013} in applying
a covariant density functional theory \cite{Ring2010} with density-dependent
couplings \cite{Lalazissis2005}. The quark phase in Set I is described by a MIT
bag model \cite{Witten1984,Haensel1986,Alcock1986,Zdunik2000}, adopting the
constant sound speed ($c_s$) parametrization \cite{Seidov1971, Zdunik2013,
  Alford2014, Alford2017}. Hence, the Set I quark phase pressure $P$ as a
function of the energy density $\epsilon$ is given by
\begin{equation}
  \label{eq:pressure-ACS}
  P(\epsilon) =
  \begin{cases}
    P_{\text{tr}}                  & \epsilon_1 \le \epsilon \le \epsilon_2\,, \\
    P_{\text{tr}} + c_s^2(\epsilon - \epsilon_2) & \epsilon \ge \epsilon_2\,,
  \end{cases}
\end{equation}
with $P_{\text{tr}}$ value of the pressure at the phase transition (which occurs
in the energy density range $\epsilon_1 \le \epsilon \le \epsilon_2$). The energy density jump is
parametrized by the value of $\xi$ defined through\footnote{In
  \cite{Paschalidis2018} the letter $j$ is used instead of $\xi$.}
\begin{equation}
  \label{eq:jump}
  \epsilon_2 = (1 + \xi) \epsilon_1\,.
\end{equation}
The values of $\epsilon_1, \xi$, $P_{\text{tr}}$ and $c_s$ that produce the Set I EOSs
(labeled here T$1$--T$9$ and TT) are listed in Table I of \cite{Paschalidis2018}.

Set II \cite{Alvarez-Castillo2019} consists of a piecewise polytropic
representation \cite{Read2009,Hebeler2013,Raithel2016} of the quark phase in
which the pressure is
\begin{equation}
  \label{eq:pressure-ACB}
  P(n) = \kappa_i \left(\frac{n}{n_0}\right)^{\Gamma_i} \quad \text{for } n_i < n < n_{i+1}\,,
\end{equation}
where $n$ is the baryon number density and $n_0$ its value at the nuclear
saturation density. The second polytrope has $\Gamma_2 = 0$ so that the pressure is
constant, corresponding to the phase transition. The various coefficients that
specify these EOSs (labeled here as A$4$--A$7$) are listed in Table II of
\cite{Paschalidis2018}. Note that EOSs with similar properties as the EOSs in
Set I and II have been obtained recently within a relativistic density
functional approach to quark
matter~\cite{Bonanno2012,Ayvazyan2013,Kaltenborn2017}.

\begin{figure}
  \centering
 \includegraphics{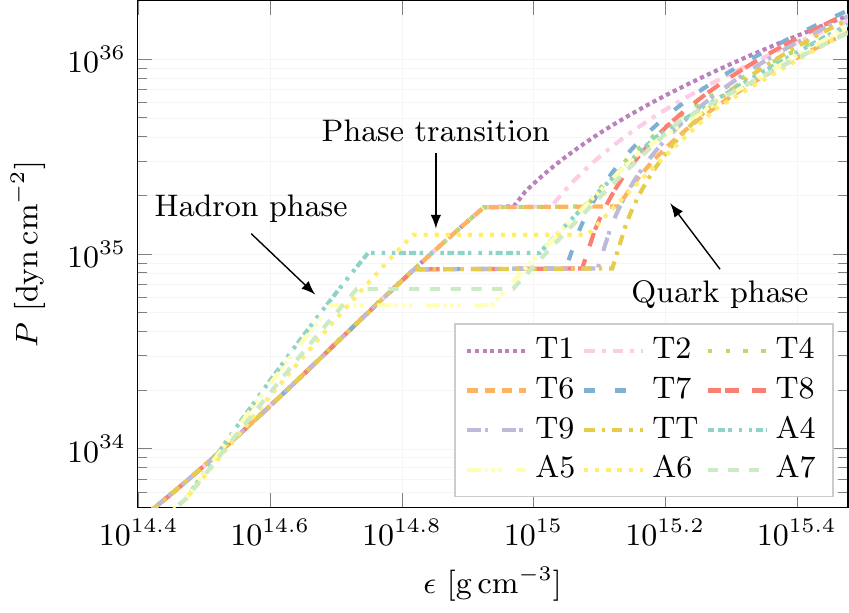}
 \caption{Pressure $P$ as a function of the energy density $\epsilon$ for the equations
   of state studied in this paper. }
  \label{fig:eos}
\end{figure}

\begin{figure}
  \centering
  \includegraphics{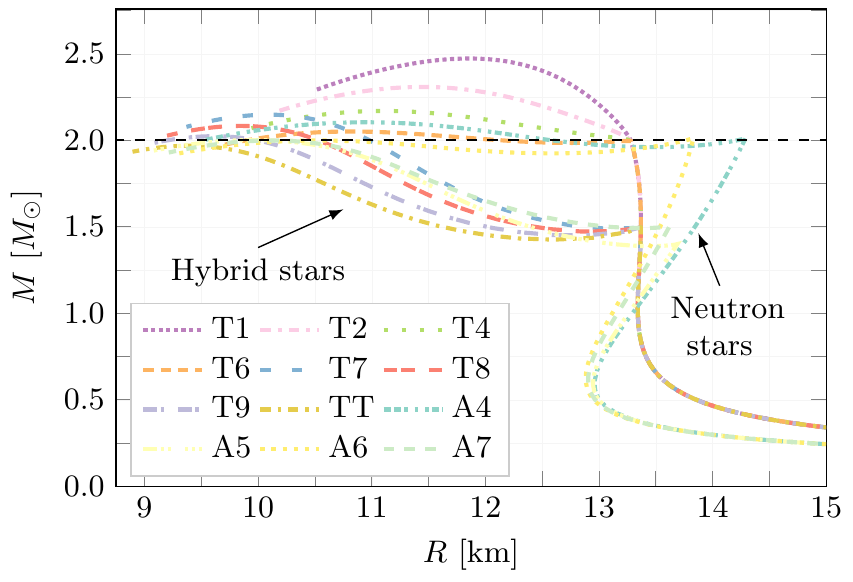}
  \caption{ Mass-radius relation for non-rotating stars with the EOSs listed in
    Table~\ref{tab:eos} and plotted in Figure~\ref{fig:eos}. Equations of state with
    quark-hadron transitions allow for a third family of relativistic stars more
    compact than neutron stars, and hence hybrid stars are on the left part of
    the curve. The hadronic and/or the hybrid branches of all the EOSs lead to
    configurations (at least marginally) compatible with the constraint of a
    $\SI{2}{\sunmass}$ pulsar (dashed line).}
  \label{fig:M-R}
\end{figure}

Figure~\ref{fig:eos} shows the pressure $P$ as a function of the energy density
$\epsilon$ for the different EOSs, and Figure~\ref{fig:M-R} depicts the resulting
mass-areal radius relationship for non-rotating compact stars. The EOSs are
compatible with the observational constraint of a pulsar mass of
$(\num{1.97}\,\pm\, \num{0.04})\,\si{\sunmass}$
\cite{Demorest2010,Antoniadis2013,Fonseca2016,Arzoumanian2018}, and half of
them are also compatible with the newly discovered
$\num{2.17}^{+0.11}_{-0.10}~\si{\sunmass}$ pulsar
J0740+6620~\cite{Cromartie2019}. The EOSs span a relatively large range of
compactness and radii ($\num{10}-\SI{13.5}{\km}$), and as such allow us to test
for universal relations. Despite the fact that some of the EOSs in our sample
are incompatible with J0740+6620, we include them in our study to check whether
existing universal relations are respected independently of the EOS.

\begin{table}
  \centering
  \caption{Labels of the equations of state employed in this paper and their
    corresponding labels in \cite{Paschalidis2018}. EOSs T$1$--TT belong to Set I
    and have a first-order phase transition with energy-density jump
    parametrized by $\xi$ as shown in Equation~\eqref{eq:jump}. Their quark phase
    is described by Equation~\eqref{eq:pressure-ACS}. On the other hand, EOSs
    A$4$--A$7$ belong to Set II, and have a quark phase parametrized as piecewise
    polytropes as shown in Equation~\eqref{eq:pressure-ACB}. We also report the
    maximum mass that a TOV star can have on the hadronic
    ($M_{\text{Max, hadronic}}^{\text{TOV}}$) and the hybrid
    ($M_{\text{Max, hybrid}}^{\text{TOV}}$) branches, respectively.}
  \label{tab:eos}
  \begin{tabular}{cc|cc}
    EOS & EOS in \cite{Paschalidis2018} & $M_{\text{Max, hadronic}}^{\text{TOV}}$ & $M_{\text{Max, hybrid}}^{\text{TOV}}$ \\
         &                                  & $[\si{\sunmass}]$          & $[\si{\sunmass}]$        \\ \hline
    T1   & ACS-I $\xi = \num{0.10}$           & \num{2.00}                 & \num{2.47}               \\
    T2   & ACS-I $\xi = \num{0.27}$           & \num{2.00}                 & \num{2.31}               \\
    T4   & ACS-I $\xi = \num{0.43}$           & \num{2.00}                 & \num{2.17}               \\
    T6   & ACS-I $\xi = \num{0.60}$           & \num{2.00}                 & \num{2.05}               \\
    T7   & ACS-II $\xi = \num{0.70}$          & \num{1.50}                 & \num{2.14}               \\
    T8   & ACS-II $\xi = \num{0.80}$          & \num{1.50}                 & \num{2.08}               \\
    T9   & ACS-II $\xi = \num{0.90}$          & \num{1.50}                 & \num{2.02}               \\
    TT   & ACS-II $\xi = \num{1.00}$          & \num{1.50}                 & \num{1.98}               \\
    A4   & ACB4                             & \num{2.01}                 & \num{2.11}               \\
    A5   & ACB5                             & \num{1.40}                 & \num{2.00}               \\
    A6   & ACB6                             & \num{2.01}                 & \num{2.00}               \\
    A7   & ACB7                             & \num{1.50}                 & \num{2.00}               \\ \hline
  \end{tabular}
\end{table}

\section{Methods}
\label{sec:methods}

We assume that neutron star matter can be described as a perfect fluid with no
meridional currents. An equilibrium model for a rotating general-relativistic
star is a stationary and axisymmetric solution of Einstein's equations coupled
with the equation of hydrostationary equilibrium. Under these assumptions, the
fluid four-velocity is given by
\begin{equation}
  \label{eq:four-velocity}
  u = (u^t, 0, 0, u^\phi)\,,
\end{equation}
and the line element ${\d s}^2 $ of the spacetime is
\begin{multline}
  \label{eq:spacetime_metric}
  {\d s}^2 = -\mathrm{e}^{\gamma + \rho} {\d t}^2 + \mathrm{e}^{\gamma - \rho}r^2 \sin^2 \theta(\d \phi
  - \omega \d t)^2 + \\
  \mathrm{e}^{2\alpha}({\d r}^2 + r^2 {\d \theta}^2)\,,
\end{multline}
with $(t, r, \theta, \phi)$ quasi-isotropic coordinates, and
$\gamma,\rho,\alpha,\omega$ spacetime potentials that depend on the coordinates
$r, \theta$ only \cite{Bardeen1971, Friedman2013, PaschalidisStergioulas2017}.

To solve the equilibrium equations, an EOS and a rotation law have to be
supplied (see~\cite{PaschalidisStergioulas2017} for a recent review on
theoretical and numerical approaches to the construction of rotating
relativistic stars). For uniform rotation, we fix $\Omega = u^\phi \slash u^t$ (the local
angular velocity of the fluid as seen by an observer at rest at infinity) to be
constant. For differential rotation, we adopt the Komatsu-Eriguchi-Hachisu (KEH)
$j$-constant law~\cite{Komatsu1989a, Komatsu1989b}, which is described by
\begin{equation}
  \label{eq:keh-law}
  u^t u_\phi = A^2 (\Omega_c - \Omega)\,,
\end{equation}
where $\Omega_c$ it is the angular velocity evaluated on the rotation axis, and
$A$ is a parameter with units of length that determines the lengthscale of
variation of angular velocity within the star. This rotation law is not the only
possible choice, but it is most commonly studied due to its simplicity (for a
summary of other differential rotation laws,
see~\cite{PaschalidisStergioulas2017}). In the following, instead of using $A$,
we work with the dimensionless parameter $\hat{A}^{-1}$ defined by
\begin{equation}
  \label{eq:Am1}
  \hat{A}^{-1} = \frac{r_e}{A}\,,
\end{equation}
where $r_e$ is the coordinate radius of the star at the equator. Stars with
$\hat{A}^{-1} = 0$ are uniformly rotating.

We use the code developed in \cite{Cook1994, Cook1994b} (the ``Cook code'') to
solve the structure equations and construct equilibrium models for rotating
stars. For a given EOS and degree of differential rotation $\hat{A}^{-1}$, a
rotating equilibrium model is built by providing the maximum energy density
$\epsilon_{\rm max}$ and the ratio of the polar ($r_p$) to equatorial coordinate radius
$r_e$. In the case of differential rotation with the KEH law, the set of
parameters $\{\epsilon_{\rm max}, r_p/r_e, \hat{A}^{-1} \}$ does not \emph{uniquely}
specify an equilibrium model, in spite of the fact that it is sufficient to
find a solution \cite{Ansorg2009}. To distinguish between physically distinct
models that have the same
$\{\epsilon_{\rm max}, r_p/r_e, \hat{A}^{-1} \}$, the parameter
\begin{equation}
  \label{eq:beta}
  \beta = -\left(\frac{r_e}{r_p}\right)^2
  \frac{\d(z^2)}{\d (\varpi^2)} \biggr\rvert_{\varpi=r_e}
\end{equation}
was introduced in \cite{Ansorg2009}. Here $\varpi=r\sin(\theta)$ and
$z=r\cos(\theta)$ are cylindrical coordinates, and the derivative is evaluated on the
surface of the star at the equator. In practice, it is more convenient to work
with a new quantity $\hat{\beta}$ defined in terms of $\beta$ as
\begin{equation}
  \label{eq:beta-hat}
\hat{\beta} = \frac{\beta}{1+\beta}\,.
\end{equation}
The parameter $\hat{\beta}$ describes how close to the mass-shedding limit a
configuration is, and approaches three limiting values according to the stellar
shape in the solution space: in the case of a spherical solution, $\beta = 1$ and so
$\hat{\beta}$ is ${1}\slash{2}$; for a model close to the mass-shedding limit, the
stellar surface is highly pinched near the equator, which implies that the
derivative appearing in Equation~\eqref{eq:beta} vanishes, and hence
$\hat{\beta}$ approaches $0$; finally, for a quasi-toroidal shape the ratio
${r_p}\slash{r_e}$ becomes very small, and hence $\beta$ becomes large. In this limit
$\hat{\beta}$ approaches $1$.

In the case of the KEH rotation law, the solution space of differentially
rotating stars is fully specifiable by the quadruplet of parameters
$\{\epsilon_{\rm max}, r_p/r_e, \hat{A}^{-1}, \hat{\beta} \}$ \cite{Ansorg2009}.
Our code does not have the capability of fixing $\hat{\beta}$ directly, but, as
in \cite{Espino2019}, we produce models with different values of $\hat{\beta}$ given
the other parameters $\{\epsilon_{\rm max}, r_p/r_e, \hat{A}^{-1}\}$ by following
specific trajectories in the parameter space. This is especially relevant for
our discussion in Appendix~\ref{sec:solut-space-diff}, where we study the
different solution types for hydrid stars.

\section{Uniform Rotation}
\label{sec:maxim-mass-supr}

In this section we focus on the case of uniform rotation. We compute the
supramassive limit mass and the maximum mass on sequences of constant angular
momentum, and test the applicability of the universal relations found
in~\cite{Breu2016} and~\cite{Bozzola2018} to the hydrid hadron-quark EOSs listed
in Table~\ref{tab:eos}. Differential rotation is treated in
Section~\ref{sec:maxim-mass-class}.

\begin{figure}
  \centering
  \includegraphics{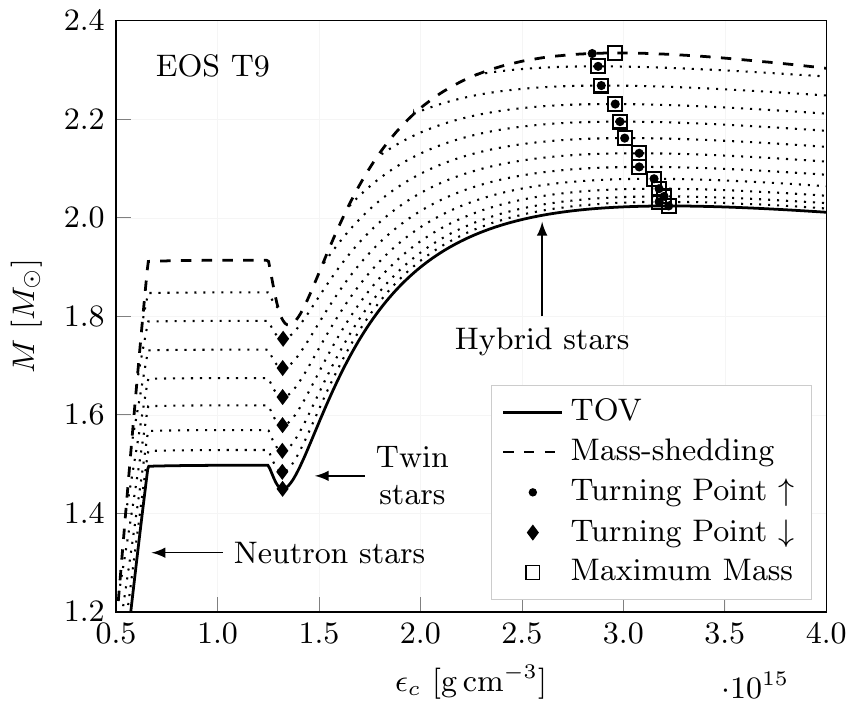}
  \caption{Constant angular momentum equilibrium sequences for uniformly
    rotating stars with EOS T9. The black solid line is the non-rotating
    sequence, and the dashed line represents the mass-shedding sequence. Black
    filled circles indicate the top turning points, whereas boxes are models
    with maximum mass along the sequence. Diamonds designate the bottom turning
    points, which indicate the least massive twin star. Most of the turning
    points lie inside a box because they are also the most massive model along
    the sequence. The flat region that occurs in the density range
    $\epsilon_c \in (\num{0.6}, \num{1.4})\,\SI{e15}{\g\per\cm\cubed}$ is a direct
    consequence of the one in Figure~\ref{fig:eos}: multiple values of
    $\epsilon_c$ lead to the same central pressure, which in turn results in models
    with the same mass.}
  \label{fig:T9}
\end{figure}

One of the most important macroscopic parameters related to an equation of state
is the maximum mass that an equilibrium rotating star can support.\footnote{The
  most massive models are typically dynamically unstable
  \cite{Friedman1988,Takami2012,Weih2018}. In this paper we ignore the issue of
  stability and focus on constructing equilibrium models. } To study
supramassive hybrid stars we construct equilibrium sequences with constant
angular momentum $J$. For these computations, we set the angular resolution of
the Cook code to 300 points, the radial to 500 and used a basis of 20 Legendre
polynomials. Then, each sequence is constructed with at least 500 models
logarithmically equispaced in the central energy density. Examples of such
sequences are depicted in Figure~\ref{fig:T9} for EOS T9, where boxes indicate
the stars with maximum mass along each $J$-constant sequence. In
Figure~\ref{fig:T9}, we also highlight other notable models: the \emph{turning
  points} indicated with filled diamonds and filled circles. In
Figure~\ref{fig:T9}, most of the circles lie inside a box, since those turning
points are also the most massive model of the sequence.

Given a sequence of equilibrium models with constant angular momentum $J$, and
varying central energy density $\epsilon_c$, the turning points are the
stationary points of the function $M(\epsilon_c)$, where $M(\epsilon_c)$ is the
gravitational mass along the sequence. For a constant $J$ sequence, a turning
point is defined by the condition
\begin{equation}
  \label{eq:turning-point-J}
  \frac{\partial M(\epsilon_c)}{\partial \epsilon_c} \biggr \rvert_J = 0\,.
\end{equation}
For constant angular momentum sequences and hadronic EOSs, the turning points
are local maxima, whereas for constant $M_0$ sequences the turning points are
local minima. By virtue of the turning point theorem
\cite{Sorkin1982,Friedman1988} that applies to uniformly rotating
configurations, the turning-point lines for $J$-constant sequences coincide with
those of $M_0$-constant sequences. In this work we located the turning points
along $J$-constant sequences by interpolating with a cubic spline around the
local maxima and minima model of the sequence, and then finding the roots of the
derivative of the cubic spline interpolant.

In the case of EOSs exhibiting a hadron-quark phase transition with a large jump
in energy density, typically each $J$-constant sequence has multiple turning
points. In this paper we focus on the two turning points of $J$-constant
sequences that occur at central energy densities higher than the
phase-transition. We call them ``top'' (labeled with an upward pointing arrow
$\uparrow$) and ``bottom'' (labeled with a downward pointing $\downarrow$),
indicating the higher and lower mass turning point, respectively. In all figures
diamonds represent bottom turning points, filled circles indicate top turning
points, and boxes designate the most massive models. Top turning-point models
are interesting because for uniformly rotating stars they detect the onset of
the instability to collapse \cite{Friedman1988, Takami2012}. On the other hand,
bottom turning points indicate the least massive twin star which must be
unstable, too. Not all EOSs in our set have bottom turning points: T1, T2 and T4
have too small a jump $\xi$ to produce such configurations (as can be seen in
Figure~\ref{fig:M-R}, where these three EOSs do not undergo a dip above the
phase transition energy density). These EOSs do not exhibit twin star solutions.

The TOV sequences are constant angular momentum ($J=0$) sequences. Their top
turning points indicate the maximum mass allowed for static
configurations,\footnote{This is not true for EOS A6, since, for this EOS,
  non-rotating stars with central energy density within the phase transition
  have mass $M = \SI{2.01}{\sunmass}$, whereas the top turning point has mass
  $M = \SI{2.00}{\sunmass}$. Since our focus is on the stable branch of the
  third family of stars, the top turning point is the maximum mass of the third
  family.} and their bottom turning points indicate the minimum mass of
non-rotating twin stars. These maximum and minimum masses provide fundamental
mass scales associated with each EOS, and will be used to normalize other
quantities. We will append the superscript ``TOV'' to designate these mass, so
that $M^{\text{TOV}}_{\text{Max}}$ and $M_{0, {\text{Max}}}^{\text{TOV}}$ are
the maximum gravitational and baryonic mass of a non-rotating hybrid star in the
third family, respectively, whereas $M^{\text{TOV}}_{\downarrow}$ and
$M_{0, \downarrow}^{\text{TOV}}$ are the minimum gravitational and rest mass for a static
twin star.\footnote{Using the maximum possible TOV mass $M = \SI{2.01}{\sunmass}$
  (instead of the maximum TOV mass in the third family
  $M = \SI{2.00}{\sunmass}$) to normalize mass scales for EOS A6 has negligible
  impact on our results.}

\subsection{The supramassive limit}
\label{sec:supramassive-limit}

\begin{figure}
  \centering
  \includegraphics{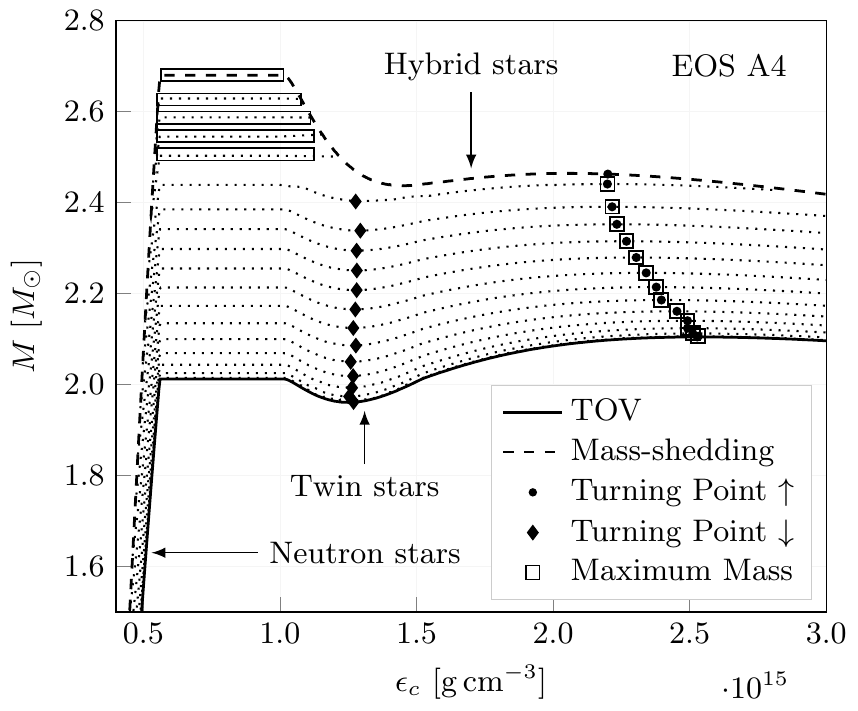}
  \caption{Same as Figure~\ref{fig:T9}, but for EOS A4. Rotation leads to a
    relative increase in the mass which is larger when the central energy
    density is in the phase-transition region. For values of $\epsilon_c$ in this
    range, a star can be spun up at higher values of the angular momentum,
    increasing the supramassive limit. The flat branches at high angular
    momentum have models with the same mass, for this reason, each of those
    stars is a maximum-mass model (empty rectangles).}
  \label{fig:TB}
\end{figure}

An EOS-insensitive relation for hadronic EOSs that relates the maximum mass
($M_{\text{Max}}$) and angular momentum along $J$-constant sequences was
recently found in \cite{Breu2016}. The resulting universal relation
(Equation~(12) in \cite{Breu2016}) is expressed in terms of $J$ normalized to
the maximum possible angular momentum $J_{\text{Max}}$ that can be achieved with
uniform rotation, and is the following
\begin{equation}
  \label{eq:breu}
  \frac{M_{\uparrow}}{M_{\text{Max}}^{\text{TOV}}} = 1 + \num{0.1316}
  \left(\frac{J}{J_{\text{Max}}} \right)^2
  + \num{0.0711} \left(\frac{J}{J_{\text{Max}}} \right)^4\,.
\end{equation}
This last equation can be used to estimate the supramassive limit for hadronic
EOSs, and yields the approximately EOS-independent value of \SI{20}{\percent} larger
than the TOV limit. Here we are interested in testing whether
Equation~\eqref{eq:breu} applies to hybrid hadron-quark EOSs. Moreover, we
explore whether a similar relation exists for hybrid EOSs not only for the top
turning points, but also for the bottom ones. As a reminder, we note that for
hybrid EOSs, the top turning point is not necessarily the most massive model
along a $J$-constant sequence, as shown for example in Figure~\ref{fig:TB}.

\subsubsection{Maximum mass}
\begin{figure}
  \centering
  \includegraphics{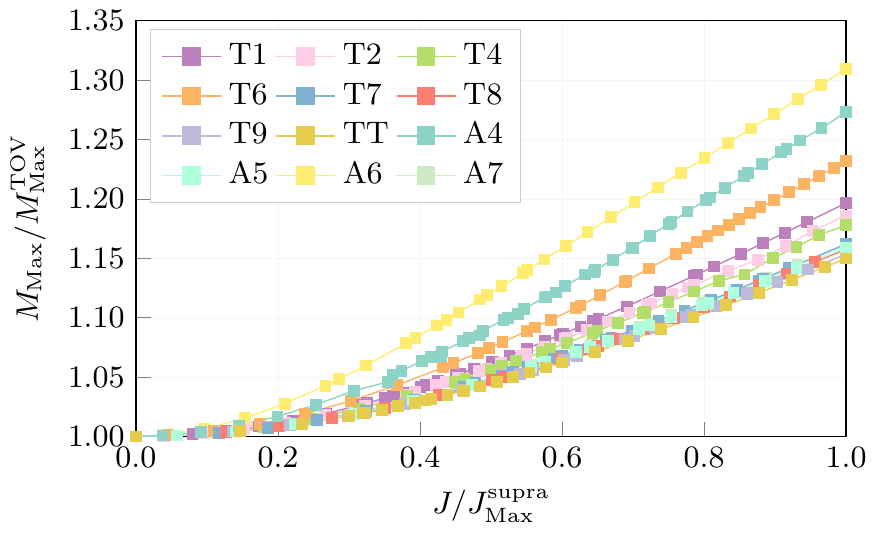}
  \caption{Maximum mass along $J$-constant sequences of uniformly rotating
    equilibrium models as a function of the angular momentum $J$ normalized to
    $J_{\text{Max}}^{\text{supra}}$ (the angular momentum of the most massive
    star on the mass-shedding sequence). There is a non-negligible dependence on
    the EOS, and the supramassive limit for EOSs in Table~\ref{tab:eos} varies
    from $\num{1.15}\,M^{\text{TOV}}_{\text{Max}}$ (for EOS TT) to
    $\num{1.31}\,M^{\text{TOV}}_{\text{Max}}$ (for EOS A6). Given this result,
    the constraints put on the EOS that employed the universality of the
    supramassive limit are valid only in the context of hadronic EOSs.}
  \label{fig:breu-max}
\end{figure}

In Figure~\ref{fig:breu-max} we plot the maximum possible gravitational mass normalized to
$M^{\text{TOV}}_{\text{Max}}$ on $J$-constant sequences as a function of $J$
normalized to $J_{\text{Max}}^{\text{supra}}$, the angular momentum of the most
massive uniformly rotating star i.e., the supramassive limit configuration. This
choice ensures that the maximum mass is reached when
$J\slash J_{\text{Max}}^{\text{supra}} = 1$. As is clear from
Figure~\ref{fig:breu-max}, some of the EOSs in our set produce uniformly
rotating stars that can support more mass than
$\num{1.20}\,M^{\text{TOV}}_{\text{Max}}$. On the other hand, some EOSs cannot
even reach such a large enhancement in mass compared to the TOV limit mass. To
be more specific, the supramassive limit mass varies from
$\num{1.15}\,M^{\text{TOV}}_{\text{Max}}$ for TT to
$\num{1.31}\,M^{\text{TOV}}_{\text{Max}}$ for A6.

\begin{figure}
  \centering
  \includegraphics{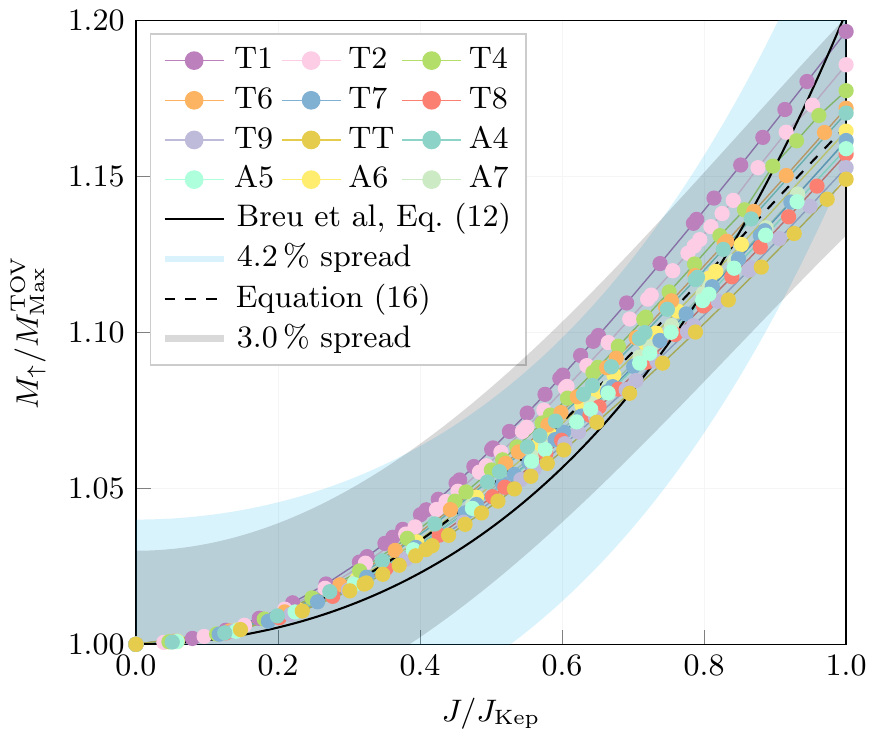}
  \caption{Normalized gravitational mass of the top turning point on
    $J$-constant sequences of uniformly rotating equilibrium models of stars in
    the third family vs the configuration's normalized angular momentum $J$. The
    gray shaded region is the $\SI{3}{\percent}$ spread. Note that in \cite{Breu2016}
    $j\slash j_{\text{Kep}}$ is used instead of $J\slash J_{\text{Kep}}$, where
    $j = J \slash M_{\text{Max}}^2$ and
    $j_{\text{Kep}} = J_{\text{Kep}} \slash M_{\text{Max}}^2$. The two quantities are
    the same, only the notation is different. For the EOSs in the Set I (see
    Section~\ref{sec:equations-state}) there is a correlation between
    $M_\uparrow\slash M^{\text{TOV}}_{\text{Max}}$ and the value of jump $\xi$: for fixed
    $J\slash J_{\text{Kep}}$, the smaller $\xi$ is, the larger
    $M_\uparrow\slash M^{\text{TOV}}_{\text{Max}}$ is. This can be clearly seen for EOSs T1,
    T2, and T4 (the first three curves from top to bottom).}
  \label{fig:breu-crit}
\end{figure}

\subsubsection{Implications for GW170817}

In~\cite{Most2018, Ruiz2018} the result that for hadronic EOSs the ratio of the
supramassive limit to the TOV limit is EOS-independent was combined with the
observation of gravitational waves from event GW170817 to set an upper bound on
the TOV limit mass. We repeat here the same argument to highlight the
differences when EOSs with a hadron-quark phase transition are considered.

As shown by simulations, the observed properties of the gamma-ray burst and the
ejecta following GW170817 suggest that the event left a rotating remnant
supported by differential rotation. The remnant then underwent delayed collapsed
to a black hole (see arguments in~\cite{Most2018, Ruiz2018}), meaning that its
mass was above the supramassive limit $M^{\text{supra}}_{\text{Max}}$. As we
already mentioned, for hadronic EOSs it is possible to write
\begin{equation}
  \label{eq:supramassive-alpha}
  M^{\text{supra}}_{\text{Max}} = \alpha M^{\text{TOV}}_{\text{Max}}\,,
\end{equation}
with $\alpha \approx \num{1.2}$, independently of the EOS. Then, assuming the low-spin
prior LIGO analysis for GW170817, the total gravitational mass of the system as
inferred by the gravitational wave signal is
$M_{\text{tot}} \approx {2.74^{+0.04}_{-0.01}}~\si{\sunmass}$~\cite{Abbott2017a}, and
since this quantity has to be greater than the supramassive limit, the following
holds
\begin{equation}
  \label{eq:gw170817-supra}
  M_{\text{tot}} \approx \SI{2.74}{\sunmass} > M^{\text{supra}}_{\text{Max}} = \alpha M^{\text{TOV}}_{\text{Max}} \,.
\end{equation}
With this inequality, the authors of~\cite{Most2018, Ruiz2018} placed a
constraint on the value of the maximum mass of a non-rotating configuration
\begin{equation}
  \label{eq:gw170817-tov}
  M^{\text{TOV}}_{\text{Max}} < \frac{M_{\text{tot}}}{\alpha} \approx \SI{2.28}{\sunmass}\,.
\end{equation}

Figure~\ref{fig:breu-max} shows the loss of EOS-independence of the supramassive
limit mass, which can be between $\SI{15}{\percent}$ to $\SI{31}{\percent}$ more
than the TOV limit mass. Assuming that we can extend
Equation~\eqref{eq:supramassive-alpha} to include hybrid stars with $\num{1.15}
\le \alpha \le \num{1.31}$, using the same steps as in the case of hadronic EOSs
the upper bound when hybrid EOSs are considered is revised to
\begin{equation}
  \label{eq:gw170817-tov-max}
  \SI{2.07}{\sunmass} \lesssim M^{\text{TOV}}_{\text{Max}} \lesssim\SI{2.38}{\sunmass}\,.
\end{equation}
The lower bound in Equation \eqref{eq:gw170817-tov-max} comes from most recent
massive pulsar J0740+6620~\cite{Cromartie2019}. The upper bound is relaxed from
$\SI{2.28}{\sunmass}$ to $\SI{2.38}{\sunmass}$ to encompass the EOSs treated in this work.

The assumption that we can extend Equation~\eqref{eq:supramassive-alpha} (with
$\num{1.15} \le \alpha \le \num{1.31}$) to include hybrid stars, should be further tested
using hybrid hadron-quark EOS with different baseline hadronic EOSs. But, this
goes beyond the scope of the current work. However, that GW170817 is compatible
with hybrid EOSs based on the fact that the supramassive limit mass should be
smaller than the inferred mass from GW170817 is demonstrated in
Table~\ref{tab:supramassive-limit}, where we show the supramassive limit mass
and the supramassive limit in the third family with each of the EOSs we treat in
this work. As is evident from the table, the largest supramassive limit
corresponds to EOS A6 which is $\SI{2.63}{\sunmass}$, and hence safely smaller
than even $\SI{2.73}{\sunmass}$ (the lower limit on the mass from GW170817). This
result lends further support to the finding of~\cite{Paschalidis2018} that
GW170817 can be interpreted as the inspiral of a binary hybrid star--neutron
star.

\begin{table}
  \centering
  \caption{Supramassive limit $M^{\text{supra}}_{\text{Max}}$ and maximum mass
    for uniformly rotating stars belonging to the third family
    $M^{\text{supra}}_{\text{Max, third}}$ for the EOSs considered in this paper
    (Table~\ref{tab:eos}). With the exception of T6, A4 and A6, for all the EOSs
    the most massive star belongs to the third family. For the other three EOSs,
    the maximum mass occurs in the phase transition (see Figure~\ref{fig:TB}).}
  \label{tab:supramassive-limit}
  \begin{tabular}{c|cccc}
    EOS & $M^{\text{supra}}_{\text{Max}}$ & \underline{$M^{\text{supra}}_{\text{Max}}$} & $M^{\text{supra}}_{\text{Max, third}}$ & \underline{$M^{\text{supra}}_{\text{Max, third}}$} \\
        & $[\si{\sunmass}]$ & $M^{\text{TOV}}_{\text{Max}}$    & $[\si{\sunmass}]$ & $M^{\text{TOV}}_{\text{Max}}$               \\ \hline
    T1 & \num{2.39} & \num{1.20} & \num{2.39} & \num{1.20} \\
    T2 & \num{2.37} & \num{1.19} & \num{2.37} & \num{1.19} \\
    T4 & \num{2.36} & \num{1.18} & \num{2.36} & \num{1.18} \\
    T6 & \num{2.46} & \num{1.23} & \num{2.34} & \num{1.17} \\
    T7 & \num{2.32} & \num{1.16} & \num{2.32} & \num{1.16} \\
    T8 & \num{2.31} & \num{1.16} & \num{2.31} & \num{1.16} \\
    T9 & \num{2.31} & \num{1.15} & \num{2.31} & \num{1.15} \\
    TT & \num{2.32} & \num{1.15} & \num{2.32} & \num{1.15} \\
    A4 & \num{2.56} & \num{1.27} & \num{2.34} & \num{1.17} \\
    A5 & \num{2.30} & \num{1.15} & \num{2.30} & \num{1.15} \\
    A6 & \num{2.63} & \num{1.31} & \num{2.33} & \num{1.16} \\
    A7 & \num{2.32} & \num{1.16} & \num{2.32} & \num{1.16} \\ \hline
  \end{tabular}
\end{table}

\subsection{Universal relations for turning points}
\label{sec:univ-relat-bott}

Turning points mark the onset of a radial instability of the star, and provide a
sufficient condition for secular axisymmetric instability in uniformly rotating
(isentropic) stars~\cite{Sorkin1982,Friedman1988, Takami2012, Weih2018} (see
also~\cite{Schiffrin:2013zta,Prabhu:2016pei}). Top turning points identify an
instability to collapse and in the case of our EOSs designate the most massive
hybrid stars. On the other hand, the bottom turning points single out the least
massive twin stars. For this reason, universal relations involving turning point
models are useful when studying the stability of rotating stars. In this
section, we focus on the turning points in the third family of compact objects
and discover a new relation, similar to Equation~\eqref{eq:breu}, for both top
and bottom turning points. We also test the universal relations reported in
\cite{Bozzola2018}, where it was shown that the angular momentum and masses of
differentially rotating turning-point models satisfy relations that are
approximately independent of either the degree of differential rotation (and
hence apply to uniformly rotating stars) or the EOS.

\subsubsection{Top turning points}

To normalize the top turning point mass and angular momentum we first define the
``Kepler turning point'' configuration in the $(\epsilon_c, M)$ plane as the model
that lies at the intersection of the mass-shedding sequence with the
turning-point line, i.e., the line connecting the top turning points along
$J$-constant sequences. Note that the Kepler turning point model is neither the
most massive nor more massive than the supramassive limit in the third family
(see, e.g, Figure~\ref{fig:TB}), but is very close to the supramassive limit of
the third family. To study universal relations of top turning points, we
normalize their angular momentum with that of the Kepler turning point, which we
denote $J_{\text{Kep}}$. In practice, we find the Kepler turning point by
least-squares fitting the turning-point line with a fifth-order polynomial, and
extending the resulting function until it intersects the mass-shedding sequence.
Given that for each EOS we construct between 15 and 40 $J$-constant sequences,
we never need to extrapolate more than $\SI{1}{\percent}$ out of the range of energy
densities where we have data.

In Figure~\ref{fig:breu-crit} we plot the top turning point mass normalized by
the TOV limit mass versus the top turning point normalized angular momentum. The
plot displays a stronger degree of universality compared to the one for the most
massive models (Figure~\ref{fig:breu-max}). The best-fitting function is
\begin{equation}
  \label{eq:new-fit-Breu}
  \frac{M_{\uparrow}}{M^{\text{TOV}}_{\text{Max}}} = 1 + \num{0.215} \left(\frac{J}{J_{\text{Kep}}}
  \right)^2 - \num{0.050} \left(\frac{J}{J_{\text{Kep}}} \right)^4\,.
\end{equation}
This expression approximates all the EOSs we study here with a spread of at most
$\SI{3}{\percent}$, which is only a little larger than the $\SI{2}{\percent}$
spread found for hadronic EOSs \cite{Breu2016}. As in the case of
\cite{Breu2016}, the spread increases with the angular momentum, and for
$J\slash J_{\text{Kep}} \lesssim \num{0.5}$ all the EOSs agree with
Equation~\eqref{eq:new-fit-Breu} to within approximately $\SI{1}{\percent}$. The
spread at higher angular momenta is mostly due to EOSs T1, T2 and T4.

By use of Equation~\eqref{eq:new-fit-Breu}, we can also find the supramassive
limit mass of the third family $M_{\text{Max, third}}^{\text{supra}}$. This is
because the flatness of the mass-shedding sequence near the supramassive limit
(see for example Figures~\ref{fig:T9} and~\ref{fig:TB}) ensures that the Kepler
turning point mass agrees with the supramassive limit mass of the third family
to better than $\SI{0.1}{\percent}$ for all EOSs we consider. Moreover, the
error in estimating the supramassive limit mass due to considering the Kepler
turning point angular momentum instead of the supramassive limit angular
momentum is smaller than the dependence from the EOS. Plugging $J =
J_{\text{Kep}}$ in Equation~\eqref{eq:new-fit-Breu}, we obtain
\begin{equation}
  \label{eq:M-max-hybrid}
  M_{\text{Max, third}}^{\text{supra}} = (\num{1.165} \pm \num{0.035}) \, M^{\text{TOV}}_{\text{Max}}\,,
\end{equation}
where the spread corresponds to the largest discrepancy from the mean value in
the set of EOSs we study (shown as the gray shaded region in
Figure~\ref{fig:breu-crit}).

In Figure~\ref{fig:breu-crit} we also compare our finding with the universal
relation of \cite{Breu2016}, since turning points were considered there, too. We
find that if we allow for a $\SI{3}{\percent}$ spread, Equation~(12) in \cite{Breu2016}
holds up to $J\slash J_{\text{Kep}} \approx \num{0.85}$, but overestimates the mass for
larger angular momenta. Equation~(12) of \cite{Breu2016} can describe the EOSs
studied here with larger angular momenta with an increased spread of
$\SI{4.2}{\percent}$ (cyan shaded region in Figure~\ref{fig:breu-crit}). Thus, our
expression~\eqref{eq:new-fit-Breu} provides a better fit to the data, and as is
clear from Figure~\ref{fig:breu-crit} Equation~\eqref{eq:new-fit-Breu} better
captures the trend of the data.

Finally, we note that for the EOSs of Set I, there is a correlation between the
jump size $\xi$ and the value of $M_\uparrow$: for a given value of
$J\slash J_{\text{Kep}}$, $M_\uparrow/M^{\rm TOV}_{\rm max}$ is larger for EOSs with smaller
$\xi$ (as is clear from the top three curves in Figure~\ref{fig:breu-crit}). As a
result, T1 has the largest normalized maximum mass
($\num{1.19}\, M^{\text{TOV}}_{\text{Max}}$) and TT the smallest
($\num{1.15}\, M^{\text{TOV}}_{\text{Max}}$). This does not happen for the EOSs
in Set II which are all comparable (regardless of the width of the phase
transition) and have no emerging trend.

Next, we consider the EOS-independent relations found in \cite{Bozzola2018}
between angular momentum, gravitational and rest mass of top turning points,
which we report here for convenience
\begin{subequations}
  \label{eq:universal-equation-bozzola}
  \begin{align}
    \frac{M_{\uparrow}}{{M}^{\text{TOV}}_{\text{Max}}}     & = 1 + 0.29
                                  \left(\frac{J}{{{M}^{\text{TOV}}_{\text{Max}}}^2}\right) ^2
                                  - 0.10
                                  \left(\frac{J}{{{M}^{\text{TOV}}_{\text{Max}}}^2}\right) ^4\,,
                                  \label{eq:J-M-b}  \\
    \frac{M_{0, \uparrow}}{{M_{0, \text{Max}}^{\text{TOV}}}} & = 1 + 0.51
                                  \left(\frac{J}{{{{M}^{\text{TOV}}_{0, \text{Max}}}}^2}\right)^2
                                  - 0.28
                                  \left(\frac{J}{{{{M}^{\text{TOV}}_{0, \text{Max}}}}^2}\right)
                                  ^4\,,
                                  \label{eq:J-M0-b} \\
                                  \frac{M_{0, \uparrow}}{{M}^{\text{TOV}}_{0, \text{Max}}} & = \num{0.93} \frac{M_{\uparrow}}{{M}^{\text{TOV}}_{\text{Max}}}
                                  - \num{0.07} \label{eq:M-M0-b}\,.
  \end{align}
\end{subequations}

\begin{figure}
  \centering
  \includegraphics{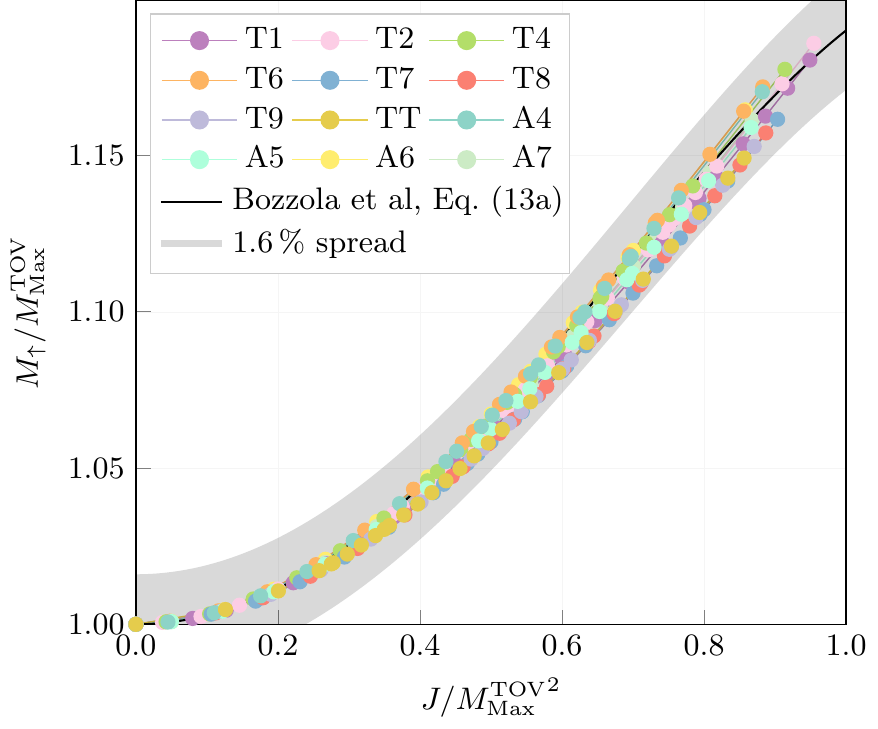}
  \caption{EOS-independent relation between angular momentum and gravitational
    mass of the top turning points for $J$-constant sequences of uniformly
    rotating stars. The solid line is Equation~(13a) in \cite{Bozzola2018}, with
    a $\SI{1.6}{\percent}$ spread indicated by the shaded region. This relation is
    satisfied by all the EOSs of Table~\ref{tab:eos}. Moreover, the relation
    provides a different parametrization of Equation~\eqref{eq:new-fit-Breu}
    with smaller spread. }
  \label{fig:J-M}
\end{figure}

Figure~\ref{fig:J-M} depicts these quantities and shows that the top turning
points are aligned with the previously known equation. In fact, we find that
these models satisfy all the three
relations~\eqref{eq:universal-equation-bozzola} with the same spread as hadronic
EOSs ($\SI{1.6}{\percent}$ for Equation~\eqref{eq:J-M-b}\footnote{This relation
  carries similar information compared to Equation~\eqref{eq:new-fit-Breu}, and exhibits
  better universality.} and
  $\SI{1.2}{\percent}$ for Equations~\eqref{eq:J-M0-b}
  and~\eqref{eq:M-M0-b}). This is a non-trivial result because in
  \cite{Bozzola2018} it was noted that purely quark stars described with the
  MIT bag model do not follow the same universal relations, even in the uniform
  rotation case. Nonetheless, here we find that for the EOSs of Set I, which
  employs a variant of the MIT bag model for the high-density phase, the
  relations are still satisfied. This suggests that the hadronic low-density
  part plays an important role in determining the universality.

\subsubsection{Bottom turning points}

We discovered that bottom turning points satisfy a relation similar to
Equation~\eqref{eq:new-fit-Breu}, if the normalizing angular momentum is that of
the Kepler bottom turning point (found at the intersection of the bottom
turning-point line with the mass shedding sequence)
\begin{equation}
  \label{eq:new-fit-Breu-bottom}
  \frac{M_{\downarrow}}{M^{\text{TOV}}_{\downarrow}} = 1 + \num{0.33}\left(\frac{J}{J_{\downarrow,\text{Kep}}}\right)^2 - \num{0.10}\left(\frac{J}{J_{\downarrow, \text{Kep}}} \right)^4\,.
\end{equation}
The spread in this equation is at most $\SI{2}{\percent}$.\footnote{Note that
T1, T2, T4 do not have any bottom turning point.}

\begin{figure}
  \centering
  \includegraphics{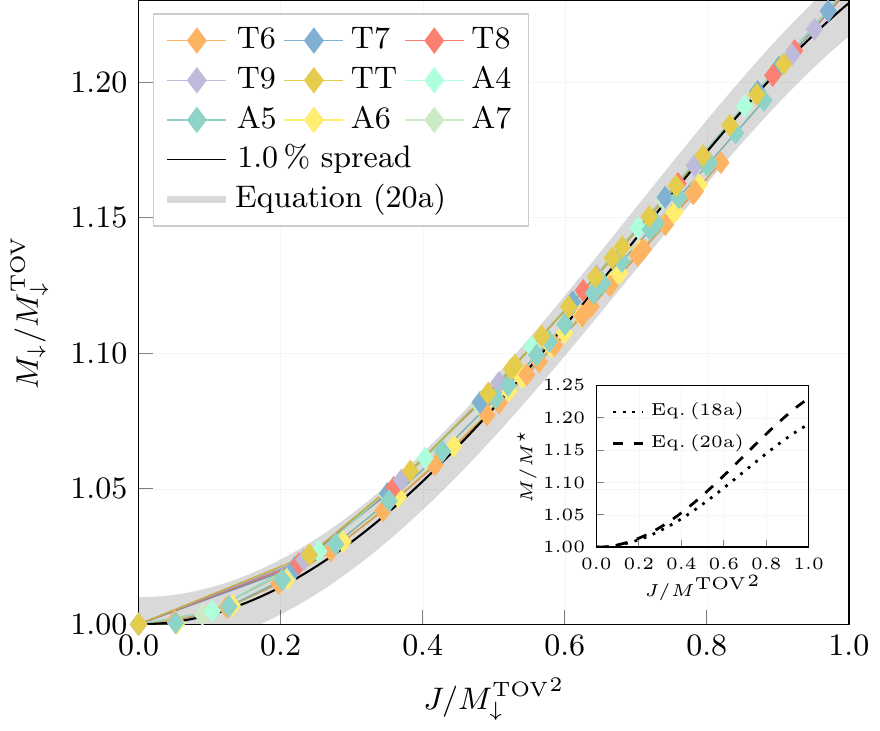}
  \caption{Same as Figure~\ref{fig:J-M}, but for bottom turning points. Here the
    solid line is Equation~\eqref{eq:J-M-dip}, and the shaded region is the
    $\SI{1}{\percent}$ spread. Bottom turning points satisfy an EOS-independent
    relation. In the inset we compare the universal relations for top and bottom
    turning points, Equations~\eqref{eq:J-M-b} and~\eqref{eq:J-M-dip}.}
  \label{fig:J-M-dip}
\end{figure}

Moreover, we find that the bottom turning points can be described with the same
Equations~\eqref{eq:universal-equation-bozzola}, but with a spread of
$\SI{3}{\percent}$. The universality becomes tighter if we consider top and bottom
turning points separately. For the bottom turning points, the best-fitting
functions are
\begin{subequations}
  \label{eq:universal-equation}
  \begin{align}
    \frac{M_{\downarrow}}{{M}^{\text{TOV}}_{\downarrow}}     & = 1 + 0.35
                                  \left(\frac{J}{{{M}^{\text{TOV}}_{\downarrow}}^2}\right) ^2
                                  - 0.12
                                  \left(\frac{J}{{{M}^{\text{TOV}}_{\downarrow}}^2}\right) ^4\,,
                                  \label{eq:J-M-dip}  \\
    \frac{M_{0, \downarrow}}{{M_{0, \downarrow}^{\text{TOV}}}} & = 1 + 0.58
                                  \left(\frac{J}{{{{M}^{\text{TOV}}_{0, \downarrow}}}^2}\right)^2
                                  - 0.35
                                  \left(\frac{J}{{{{M}^{\text{TOV}}_{0, \downarrow}}}^2}\right)
                                  ^4\,,
                                  \label{eq:J-M0-dip} \\
    \frac{M_{0, \downarrow}}{{M}^{\text{TOV}}_{0, \downarrow}} & = \num{1.016} \frac{M_{\downarrow}}{{M}^{\text{TOV}}_{\downarrow}} -
                                  \num{0.016} \label{eq:M-M0-dip}\,,
  \end{align}
\end{subequations}
with largest spread of $\SI{1}{\percent}$ for the first two and
$\SI{0.2}{\percent}$ for the last. Figure~\ref{fig:J-M-dip} shows
Equation~\eqref{eq:J-M-dip}, and in the inset we compare it to
Equation~\eqref{eq:J-M-b}, which is the corresponding universal relation for top
turning points. Among the three universal relations,
Equations~\eqref{eq:J-M-dip} and~\eqref{eq:J-M-b} are the ones in which top and
bottom turning point differ the most.

\section{Differential Rotation}
\label{sec:maxim-mass-class}

In this section we consider differentially rotating stars with hybrid EOSs and
compute the maximum mass that can be supported. In this context, it is more
common to consider the rest mass instead of the gravitational mass
\cite{Morrison2004}. Nevertheless, the results that will follow are
qualitatively the same also for the gravitational mass.

\begin{figure*}
  \centering
 \includegraphics{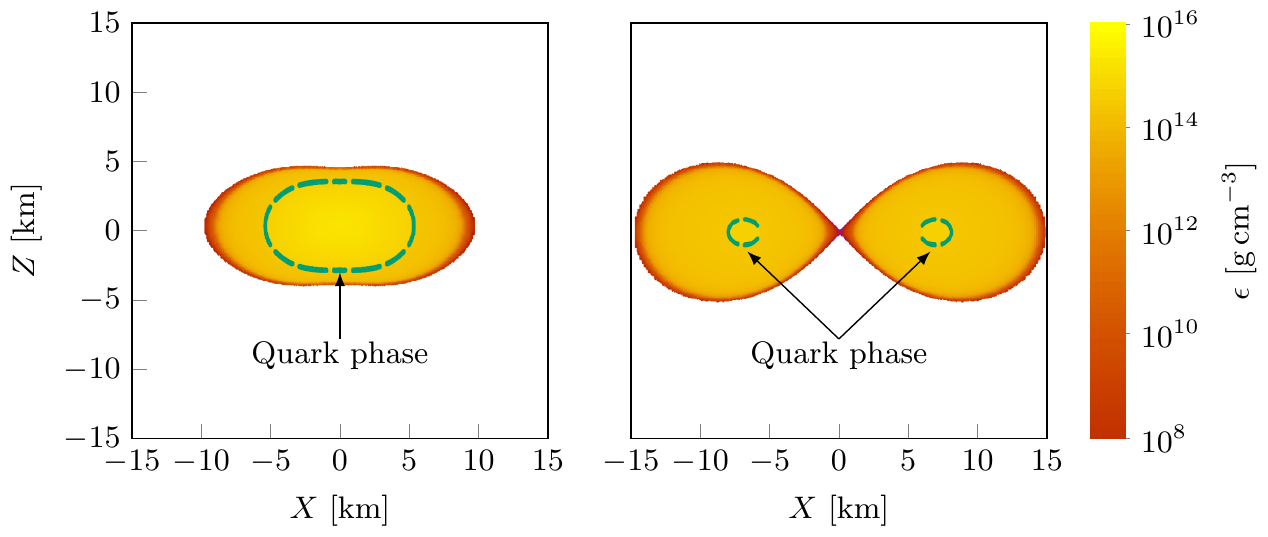}
 \caption{Examples of meridional contours of the energy density for spheroidal
   (left) and quasi-toroidal (right) stars for EOS A5. The green dashed line
   locates the quark core by outlining the contour where the energy density
   equals the value of energy density at the onset of the phase transition.}
  \label{fig:merid-A5}
\end{figure*}

The KEH rotation law of Equation~\eqref{eq:keh-law} produces stars with two
different topologies: spheroidal and quasi-toroidal. In quasi-toroidal
configurations the maximum energy density occurs in a ring around the stellar
geometric center, whereas in spheroidal stars the location of the maximum energy
density coincides with geometric center of the star. Examples of such
configurations are reported in Figure~\ref{fig:merid-A5}, where we show
meridional energy density contours for a spheroidal and a quasi-toroidal
solution. In this paper, we focus primarily on spheroidal models because
quasi-toroidal configurations are likely to be unstable \cite{Espino2019b}.

\subsection{Maximum mass}
\label{sec:maximum-mass}

To find the maximum rest mass for each EOS, we fix the degree of differential
rotation $\hat{A}^{-1}$, and scan the maximum energy density in the range
$\epsilon_{\text{max}} \in (\num{0.5}, \num{3})~\SI{e15}{\g\per\cm\cubed}$, which is
where the most massive models always lie for the EOSs considered here. For each
value of $\epsilon_{\text{max}}$ we construct equilibrium models with decreasing value
of $r_p \slash r_e$ until we find a quasi-toroidal solution or we reach the
mass-shedding limit. The last configuration built before the quasi-toroidal or
mass-shedding limit is reached is the maximum mass for that value of
$\epsilon_{\text{max}}$ in our search.

\begin{figure}
  \centering
\includegraphics{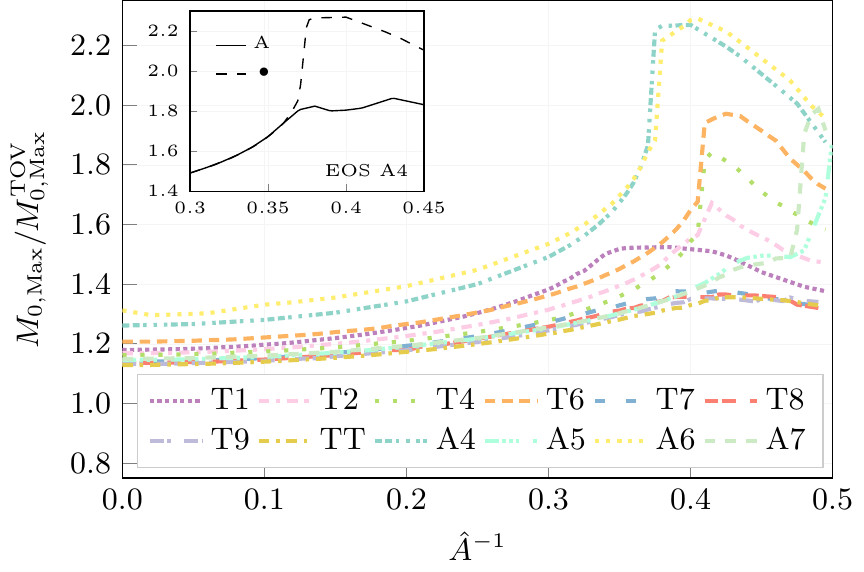}
\caption{Maximum rest mass normalized to the TOV limit rest mass
  $M_{0,\rm Max}/M_{0, {\text{Max}}}^{\rm TOV}$ as a function of degree of
  differential rotation $\hat{A}^{-1}$ for each EOS in our study
  (Table~\ref{tab:eos}). We only show the curves relevant for the maximum rest
  mass of spheroidal stars. In the inset we show the increase in mass for EOS A4
  including all spheroidal stars and only type A stars (see
  Appendix~\ref{sec:solut-space-diff}).}
  \label{fig:M0max-Am1}
\end{figure}

In Figure~\ref{fig:M0max-Am1} we show the maximum rest mass normalized to the
rest-mass of the TOV limit as a function of $\hat{A}^{-1}$. The behavior is
similar to that seen for polytropes \cite{Studzinska2016} and hadronic EOSs
\cite{Espino2019} and it is followed very closely by the gravitational mass: at
relatively low values of $\hat{A}^{-1}$ ($\lesssim 0.25$), corresponding to rotational
profiles closer to uniform rotation, there are only modest increases in the rest
mass, not exceeding $\SI{40}{\percent}$. For higher values of $\hat{A}^{-1}$, the
increase of $M_{0,\text{Max}}$ compared to the TOV limit grows until it reaches
a maximum and then begins to decrease again as $\hat{A}^{-1}$ is increased
further. The largest gain in rest mass is seen for EOS A6 ($\SI{123}{\percent}$ more
than the TOV limit), which is also the EOS with the highest gain in the
supramassive limit (see Section~\ref{sec:maxim-mass-supr}).

In Table~\ref{tab:maxmass_SPH} we report relevant properties of the most massive
spheroidal models found in our search of the parameter space. Larger increases
in the rest mass are possible if we consider quasi-toroidal stars and possibly
other differential rotation laws. Thus, these masses should be considered as
lower-limits on the maximum mass of even spheroidal stars. Moreover, finding the
``absolute'' maximum rest mass model for hybrid EOSs with the KEH law depends
sensitively on the resolution in $\epsilon_{\rm max}$, $r_p/r_e$, and
$\hat{A}^{-1}$ used to scan the parameter space. For the models presented in
Table \ref{tab:maxmass_SPH} we use a step in $\epsilon_{\rm max}$ and $r_p/r_e$
of 0.01, and a step in $\hat{A}^{-1}$ of 0.005. Even with this high resolution
in $\hat{A}^{-1}$, we found large increases in the rest mass over small ranges
of $\hat{A}^{-1}$. Such increases in mass correspond to a transition to a
different \emph{solution type}, as shown in the inset of
Figure~\ref{fig:M0max-Am1}, where we show the maximum rest mass type A models
along with the maximum rest mass spheroidal solutions (which also include type C
models) for the A4 EOS. We explore the maximum rest mass of each solution type
in Appendix~\ref{sec:solut-space-diff}, where more details on the different
types of solutions of differentially rotating stars are also presented.

\begin{table*}[htp]
  \centering
  \caption{Maximum rest mass spheroidal models for the EOSs considered in this
    work. Shown are the values of the degree of differential rotation
    $\hat{A}^{-1}$, the maximum energy density $\epsilon_{\rm max}$ in units of
    $\SI{e15}{\g\per\cm\cubed}$, the ratio of polar to equatorial radius
    ${r_p}\slash{r_e}$, the ratio of kinetic to gravitational potential energy
    ${T}\slash{\lvert W \rvert}$, the ratio of central to equatorial angular velocity
    ${\Omega_c}\slash{\Omega_e}$, the circumferential radius at the equator
    $R_e$ in units of $\si{\km}$, the dimensionless spin ${J}\slash{M^2}$, the mass
    $M_{\rm ADM}$ in units of $\si{\sunmass}$, the ratio of the mass
    $M_{\rm ADM}$ to the TOV limit mass $M_{\text{Max}}^{\text{TOV}}$, the rest
    mass $M_0$ in units of $\si{\sunmass}$ and the ratio of the rest mass to TOV
    limit rest mass $M_{0, \text{Max}}^{\text{TOV}}$. The increase in maximum
    mass compared to the non-rotating limit is between $\sim \SI{40}{\percent}$ and
    $\sim \SI{123}{\percent}$. Quantities of particular interest are
    $T/\lvert W\rvert$ that can be larger than $\sim \num{0.26}$, which is the
    threshold for the dynamical bar mode instability (see
    e.g.~\cite{PaschalidisStergioulas2017} for a review), and the rest mass
    which can be larger than twice the TOV limit -- such configurations are
    classified as ubermassive according to~\cite{Espino2019}.}
  \label{tab:maxmass_SPH}

\begin{tabular}{c|cccccccccccc}
EOS & $\hat{A}^{-1}$ & $\epsilon_{\rm max}$ & \underline{$r_p$}    & \underline{\hspace{4pt}$T_{}$\hspace{4pt}} & \underline{$\Omega_c$}   & $R_e$          & \underline{\hspace{3pt}$J_{}$\hspace{3pt}}   & $M$           & \underline{\hspace{6pt}$M_{}$\hspace{6pt}} & $M_0$         & \underline{\hspace{6pt}$M_0$\hspace{6pt}}         \\
&                  & $[\SI{e15}{\g\per\cm\cubed}]$ &   $ r_e$         &     $\lvert W \rvert$          &   $\Omega_e$          & $[\si{\km}]$ &    $M^2$         & $[\si{\sunmass}]$ & $M_{\text{Max}}^{\rm TOV}$ & $[\si{\sunmass}]$ & $M_{0, \text{Max}}^{\rm TOV}$        \\ \hline
T1 & $\num{0.390}$ & $\num{1.03}$ & $\num{0.35}$ & $\num{0.265}$ & $\num{1.930}$ & $\num{18.25}$ &
 $\num{0.922}$ & $\num{3.860}$ & $\num{1.563}$ & $\num{4.526}$ & $\num{1.542}$ \\
T2 & $\num{0.410}$ & $\num{1.09}$ & $\num{0.32}$ & $\num{0.276}$ & $\num{1.965}$ & $\num{19.35}$ &
 $\num{0.943}$ & $\num{3.974}$ & $\num{1.720}$ & $\num{4.662}$ & $\num{1.699}$ \\
T4 & $\num{0.415}$ & $\num{1.23}$ & $\num{0.31}$ & $\num{0.280}$ & $\num{1.990}$ & $\num{19.53}$ &
 $\num{0.948}$ & $\num{4.008}$ & $\num{1.847}$ & $\num{4.699}$ & $\num{1.842}$ \\
T6 & $\num{0.425}$ & $\num{1.38}$ & $\num{0.30}$ & $\num{0.283}$ & $\num{2.038}$ & $\num{19.58}$ &
 $\num{0.952}$ & $\num{4.018}$ & $\num{2.004}$ & $\num{4.705}$ & $\num{1.971}$ \\
T7 & $\num{0.415}$ & $\num{1.71}$ & $\num{0.36}$ & $\num{0.247}$ & $\num{2.064}$ & $\num{14.70}$ &
 $\num{0.892}$ & $\num{3.097}$ & $\num{1.447}$ & $\num{3.633}$ & $\num{1.393}$ \\
T8 & $\num{0.395}$ & $\num{2.07}$ & $\num{0.38}$ & $\num{0.238}$ & $\num{2.040}$ & $\num{13.55}$ &
 $\num{0.878}$ & $\num{2.944}$ & $\num{1.415}$ & $\num{3.450}$ & $\num{1.368}$ \\
T9 & $\num{0.420}$ & $\num{2.08}$ & $\num{0.38}$ & $\num{0.235}$ & $\num{2.079}$ & $\num{13.49}$ &
 $\num{0.875}$ & $\num{2.845}$ & $\num{1.408}$ & $\num{3.325}$ & $\num{1.361}$ \\
TT & $\num{0.430}$ & $\num{2.22}$ & $\num{0.38}$ & $\num{0.234}$ & $\num{2.116}$ & $\num{13.11}$ &
 $\num{0.872}$ & $\num{2.758}$ & $\num{1.393}$ & $\num{3.216}$ & $\num{1.356}$ \\
A4 & $\num{0.405}$ & $\num{1.11}$ & $\num{0.26}$ & $\num{0.304}$ & $\num{1.985}$ & $\num{22.69}$ &
 $\num{0.985}$ & $\num{4.688}$ & $\num{2.222}$ & $\num{5.593}$ & $\num{2.275}$ \\
A5 & $\num{0.510}$ & $\num{0.89}$ & $\num{0.28}$ & $\num{0.293}$ & $\num{1.990}$ & $\num{23.07}$ &
 $\num{1.013}$ & $\num{3.951}$ & $\num{1.976}$ & $\num{4.577}$ & $\num{1.937}$ \\
A6 & $\num{0.410}$ & $\num{1.31}$ & $\num{0.25}$ & $\num{0.305}$ & $\num{2.036}$ & $\num{22.00}$ &
 $\num{0.983}$ & $\num{4.557}$ & $\num{2.279}$ & $\num{5.431}$ & $\num{2.334}$ \\
A7 & $\num{0.490}$ & $\num{0.96}$ & $\num{0.28}$ & $\num{0.292}$ & $\num{1.992}$ & $\num{22.44}$ &
 $\num{1.000}$ & $\num{4.013}$ & $\num{2.007}$ & $\num{4.675}$ & $\num{1.987}$ \\ \hline
 \end{tabular}
\end{table*}

\subsection{Universal relations}
\label{sec:universal-relations-1}

In \cite{Bozzola2018} it was found that for hadronic EOSs and restricting to
spheroidal models, Equation~\eqref{eq:universal-equation-bozzola} applies not
only to uniformly rotating stars, but also to differentially rotating stars,
and, what is more, the relation is approximately insensitive to the degree of
differential rotation (with maximum spread of about $\SI{1}{\percent}$ even for high
degrees of differential rotation). Moreover, recently, in \cite{Zhou:2019hyy} it
was noted that for purely strange quark-matter stars the relations between
angular momentum and masses of turning points follow equations approximately
independent of $\hat{A}^{-1}$ (with spread below $\SI{2}{\percent}$), although the
relation is not the same as for hadronic EOSs \cite{Bozzola2018}.
Here, we construct sequences of differentially rotating equilibrium models with
constant angular momentum and various degrees of differential rotation in the
range $\hat{A}^{-1} \in (\num{0}, \num{2})$ to test whether this property holds
for the hybrid hadron-quark EOSs we treat in this work.

We find that relations~\eqref{eq:universal-equation-bozzola} apply approximately
to hybrid EOSs for degrees of differential rotation
$\hat{A}^{-1} \in (\num{0}, \num{2})$, but with a larger spread of about
$\SI{3}{\percent}$ and with an evident (albeit weak) dependence on
$\hat{A}^{-1}$. Furthermore, the larger the degree of differential rotation, the
larger the spread becomes. For this reason, the universality found in
\cite{Bozzola2018} appears to be broken in the case of hybrid stars.

\begin{figure}
  \centering
  \includegraphics{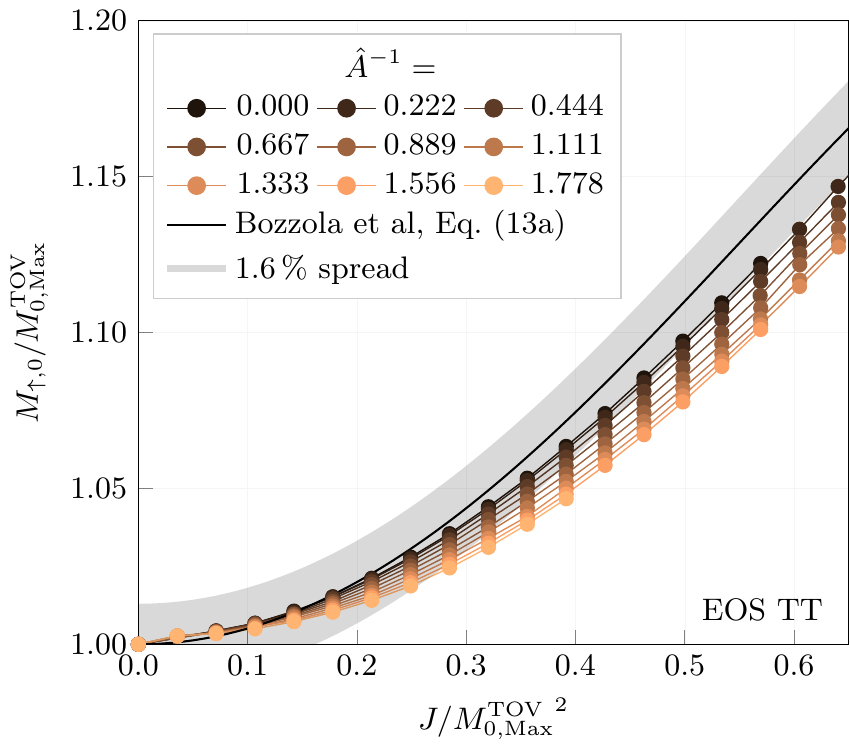}
  \caption{Normalized rest mass as a function of the normalized angular momentum
    for top turning points on $J$-constant sequences with various degrees of
    differential rotation. The plot corresponds to EOS TT. The turning points
    are compatible with the previously known relation only in the case of
    uniform rotation, but there is a clear dependence on the degree of
    differential rotation. This is the case where the violation of the universal
    relation satisfied by the uniformly rotating case is the most evident, and
    shows that the relations found in \cite{Bozzola2018} cannot be considered
    $\hat{A}^{-1}$-independent for the EOSs considered in this work.}
  \label{fig:J-M0-diff}
\end{figure}

\begin{figure}
  \includegraphics{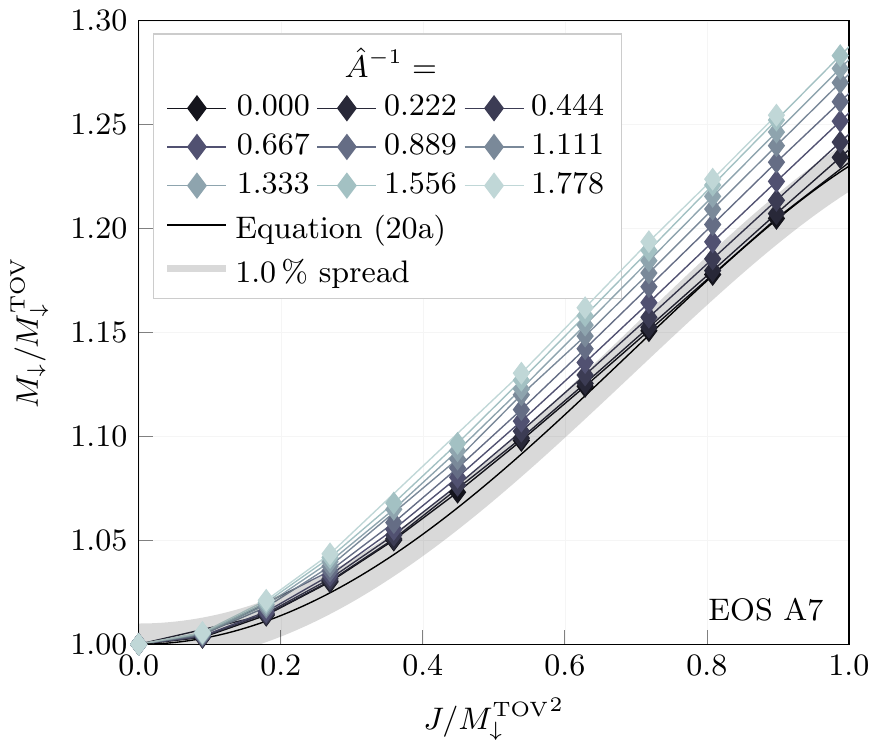}
  \caption{Similar as Figure~\ref{fig:J-M0-diff}, but for rest mass and angular
    momentum of bottom turning points, and for EOS A7. In this case, too, there
    is a clear dependence on the degree of differential rotation, and so
    Equations~\eqref{eq:universal-equation} are not $\hat{A}^{-1}$-independent.
    Among the EOS considered in this study, this is the EOS with the strongest
    violation of the universal behavior.}
  \label{fig:J-M-diff-dip}
\end{figure}

For top turning points, we show an example of the above conclusions in
Figure~\ref{fig:J-M0-diff}, where we report the results from $J$-constant
sequences constructed with EOS TT. We found that there is a correlation between
the jump size $\xi$ and the loss of the universality: the larger the jump, the
stronger the deviation from universality. Hence, the example in
Figure~\ref{fig:J-M0-diff} is one of the cases that violates the universality
the strongest. For most of the EOSs we treat, the bottom turning point universal
relations exhibit a smaller spread with increasing $\hat{A}^{-1}$ (below
$\SI{2}{\percent}$), but they still violate Equations~\eqref{eq:universal-equation} for
some EOSs, as reported in Figure~\ref{fig:J-M-diff-dip} for EOS A7. For this
EOS, the deviation from Equations~\eqref{eq:universal-equation} can be up to
$\SI{4}{\percent}$.

Both Figure~\ref{fig:J-M0-diff} and~\ref{fig:J-M-diff-dip} show that there is a
clear dependence on $\hat{A}^{-1}$. Hence, we conclude that the universal
relations found in \cite{Bozzola2018} are EOS-independent for uniform rotation,
but not for differential rotation, as they acquire a dependence on
$\hat{A}^{-1}$ which is not the same for all the EOSs. The regularity and
smoothness of sequences in the figures suggests that it is possible to
parametrize the dependence on $\hat{A}^{-1}$ using some scale characteristic of
the EOS. However, exploring this goes beyond the scope of the current work.

Finally, we point out that for sufficiently high values of $\hat{A}^{-1}$ and
$J$, in some EOSs the $J$-constant sequences exhibit no bottom turning points.
This is because stars whose central energy density is in the phase transition
region become more massive when spun up compared to the ones with higher central
energy density. This means that $J$-constant sequences in the $(\epsilon_c, M)$ plane
tend to become flatter and have shallower dips (see, for example,
Figure~\ref{fig:TB}). In some other cases with high degree of differential
rotation and angular momentum we could not even identify the top turning point,
because the change of concavity is not resolved by the accuracy of our code.

\section{Conclusions}
\label{sec:conclusions}

The equation of state of nuclear matter is uncertain at densities larger than
nuclear saturation density $\rho_0$. Some of the proposed EOS models undergo a
phase transition from hadronic to quark matter which can result in hadron-quark
hybrid stars. For EOSs with a sufficiently large jump in energy density over
which the pressure remains constant, a third family of compact objects emerges.
In this work we analyzed the properties of rotating relativistic stars using
EOSs with first-order hadron-quark phase transitions for many of which a third
family of stars emerges. To be more specific, we employed the quark matter
parametrizations introduced in \cite{Paschalidis2018} for the EOSs developed in
\cite{Alford2017} and \cite{Alvarez-Castillo2019}.

We employed the Cook code \cite{Cook1994, Cook1994b} to build rotating
relativistic stars and we studied equilibrium sequences with constant angular
momentum and varying central energy density. We found that the maximum mass a
uniformly rotating hybrid star can support (the so-called supramassive limit) is
not an EOS-independent quantity, but it varies from $\SI{15}{\percent}$ to
$\SI{31}{\percent}$ more than the TOV limit mass, in contrast to previous work
\cite{Lasota1996, Breu2016} which highlighted an approximately EOS-independent
increase of about $\SI{20}{\percent}$ for hadronic EOSs. This implies that some
constraints placed on the equation of state based on the universality of the
supramassive limit and GW170817 \cite{Most2018,Ruiz2018} do not apply to EOSs
like the ones considered here. However, the supramassive limit mass of the EOSs
we adopt is consistent with GW170817, providing further support that hybrid
hadron-quark EOSs can describe GW170817.

We located the turning points on constant angular momentum sequences, which are
identified by the stationarity condition of Equation~\eqref{eq:turning-point-J}.
We defined the bottom and top turning points along a $J$-constant sequence as
those that correspond to the least massive twin star and the most massive star
in the third family, respectively. Top turning points satisfy a universal
relation [Equation~\eqref{eq:new-fit-Breu}] with spread of about
$\SI{3}{\percent}$. With this relation, we found that the supramassive limit of the
third family is given by
\begin{equation}
  \label{eq:M-max-hybrid-conc}
  M_{\text{Max, third}}^{\text{supra}} = (\num{1.165} \pm \num{0.035}) \, M^{\text{TOV}}_{\text{Max}}\,.
\end{equation}

Enabling differential rotation and focusing on spheroidal models, we found that
the maximum mass (both baryonic and gravitational) can increase by as much as
$\SI{123}{\percent}$ compared to the non-rotating case. This enhancement compared to
the TOV limit mass, represents a lower limit of what can be achieved with
differential rotation.

Finally, we investigated the applicability of universal relations for turning
points found in \cite{Bozzola2018}, who reported relations between the angular
momentum, and the masses of turning points that are independent of the EOS and
the degree of differential rotation. We found that the relations in
\cite{Bozzola2018} hold for the top turning points for uniformly rotating stars.
We also discovered similar universal relations for the bottom turning points for
uniformly rotating stars. However, both universal relations are violated when
differential rotation is enabled.

In future work we will consider constant rest-mass sequences, and explore
additional hybrid equations of state with varying baseline hadronic EOSs, to
further test some of the results we reported in this work.

\section*{Acknowledgments}

We are grateful to S.\ L.\ Shapiro for access to the code that we used to build
equilibrium models for rotating relativistic stars, and to D.\ Alvarez-Castillo,
D.\ Blanschke and A.\ Sedrakian for giving us permission to use the equations of
state developed in~\cite{Paschalidis2018}. We also thank N.\ Stergioulas for
access to the \texttt{RNS} code \cite{Stergioulas1995, Nozawa1998} that was
used to cross-check some of the results presented here. Computations were in
part performed on the Ocelote cluster at The University of Arizona. This
research used also resources provided by the Open Science Grid
\cite{Pordes2007}, which is supported by the National Science Foundation and
the U.S.\ Department of Energy's Office of Science.

\appendix
\section{Appendix: Solution Space of Differentially Rotating Stars}
\label{sec:solut-space-diff}

In this appendix we discuss the solution space of equilibrium models for stars
rotating with the KEH law of Equation~\eqref{eq:keh-law}. We find features in
the solution space which are consistent with both hadronic and strange quark
star EOSs, while also finding others which are unique to hybrid EOSs. In the
case of differential rotation with the KEH law, solutions for a given maximum
energy density can be separated on the $(r_p/r_e,\hat\beta)$ plane into four classes
depending on the degree of differential rotation. These classes are known as
types A, B, C and D \cite{Ansorg2009}. The first, type A, consists of spheroidal
configurations with a relatively low degree of differential rotation. All
uniformly rotating stars, if seen as differentially rotating with
$\hat{A}^{-1} = 0$, belong to this family. Type B solutions are characterized by
a high central angular velocity but relatively low degree of differential
rotation. These models are quasi-toroidal, since the maximum energy density does
not occur in the geometric center of the star. In Figure~\ref{fig:merid-A5}, we
show a comparison between the energy density profiles of a type A and a type B
solution. The third category (type C) contains spheroids and quasi-toroids with
higher degrees of differential rotation than types A and B. Finally, type D
stars are typically highly pinched at the equator and quasi-toroidal. We were
unable to construct type D solutions with the Cook code, but this is not a
severe limitation on our study since these configurations are not likely to be
found in Nature because their sequences start from a mass-shedding limit and end
at a mass-shedding limit \cite{Studzinska2016}.

\begin{figure}
  \centering
  \includegraphics{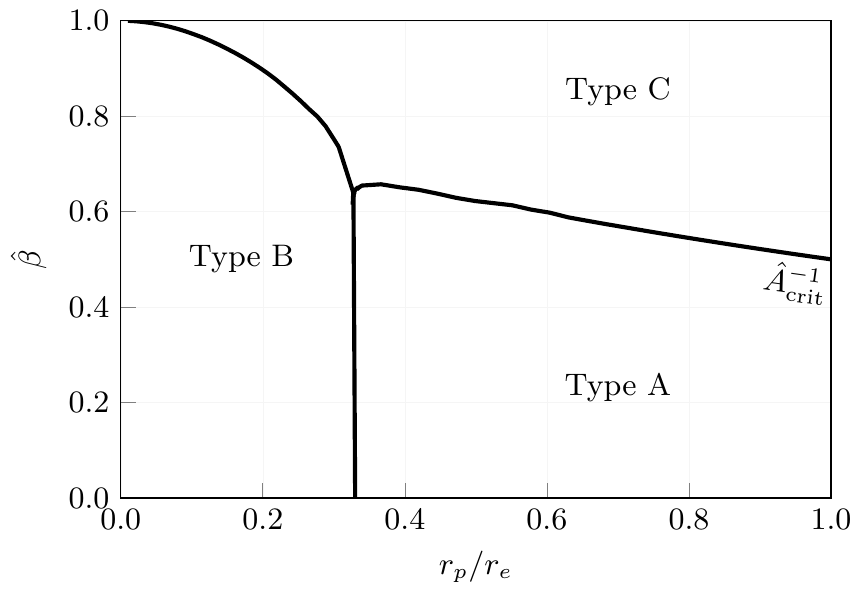}
  \caption{Mass-shedding parameter $\hat{\beta}$, calculated using
    Equation~\eqref{eq:beta-hat}, as a function of the ratio of polar to
    equatorial radii $r_p/r_e$ for a fixed energy density of
    $\epsilon_{\rm max} = \SI{1.2e15}{\g\per\cm\cubed}$. The bold black line
    corresponds to the critical degree of differential rotation
    $\hat{A}^{-1}_{\rm crit}= \num{0.611}$, defined by the saddle point on the
    $(r_p/r_e,\hat\beta)$ plane at which all solution types we are able to construct
    with the Cook code co-exist for this value of the energy density (see
    Figure~2 in \cite{Ansorg2009} for a complete plot for polytropes). The Cook
    code is not able to produce type D stars, which in this plot would reside at
    values of $r_p \slash r_e \lesssim \num{0.33}$ (the vertical leg of the separatrix), and
    $0 \leq \hat{\beta} \lesssim \num{0.6}$ \cite{Ansorg2009}.}
  \label{fig:Beta-rp}
\end{figure}

\begin{figure}
  \centering
 \includegraphics{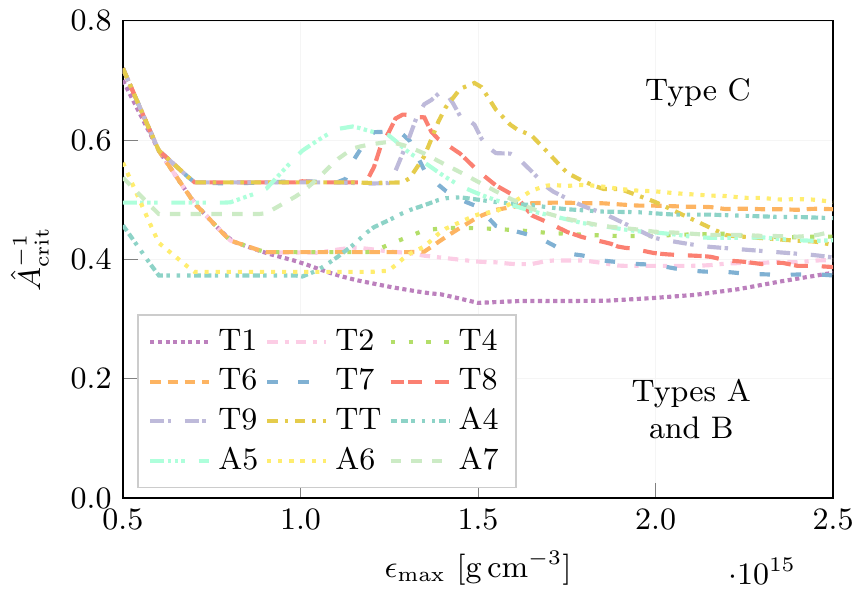}
 \caption{Critical value of the degree of differential rotation $\hat{A}^{-1}$
   as a function of the maximum energy density $\epsilon_{\rm max}$ for the EOSs in
   this study. Models of a given value of $\epsilon_{\rm max}$ with
   $\hat{A}^{-1} < \hat{A}^{-1}_{\rm crit}$ are type A or B models, and
   solutions with $\hat{A}^{-1} > \hat{A}^{-1}_{\rm crit}$ are type C. For some
   values of $\hat{A}^{-1}$, equilibrium sequences with varying maximum energy
   density change the type along the sequence.}
  \label{fig:separatrix}
\end{figure}

\begin{table*}[htp]
  \centering
  \caption{Maximum rest mass differentially rotating models for the
    T1, T8, and A6 EOSs. Shown are the solution types (A, B, or C),
    the value of the degree of differential rotation $\hat{A}^{-1}$,
    the maximum energy density $\epsilon_{\rm max}$ in units of
    $\SI{e15}{\g\per\cm\cubed}$, the ratio of polar to equatorial
    radius ${r_p}\slash{r_e}$, the ratio of kinetic to gravitational
    potential energy ${T}\slash{\lvert W \rvert}$, the ratio of
    central to equatorial angular velocity
    ${\Omega_c}\slash{\Omega_e}$, the circumferential radius at the
    equator $R_e$ in units of $\si{\km}$, the dimensionless spin
    ${J}\slash{M^2}$, the ADM mass $M$ in units of $\si{\sunmass}$,
    the ratio of the mass to the TOV limit mass, the rest mass
    in units of $\si{\sunmass}$ and the ratio of the rest mass to TOV
    limit rest mass $M_{0, \text{Max}}^{\text{TOV}}$. }
  \label{tab:maxmass_ABC}
  \begin{tabular}{cc|cccccccccccc}
    EOS & Type & $\hat{A}^{-1}$ & $\epsilon_{\rm max}$ & \underline{$r_p$}    & \underline{\hspace{4pt}$T_{}$\hspace{4pt}} & \underline{$\Omega_c$}   & $R_e$          & \underline{\hspace{3pt}$J_{}$\hspace{3pt}}   & $M$           & \underline{\hspace{6pt}$M_{}$\hspace{6pt}} & $M_0$         & \underline{\hspace{6pt}$M_0$\hspace{6pt}}         \\
 &  &  & $[\SI{e15}{\g\per\cm\cubed}]$ & $r_e$    & $\lvert W \rvert$ & $\Omega_e$   & $[\si{\km}]$          & $M^2$   & $[\si{\sunmass}]$  & $M_{\text{Max}}^{\rm TOV}$ & $[\si{\sunmass}]$   & $M_{0, \text{Max}}^{\rm TOV}$         \\
\hline
    T1 & A & $\num{0.35}$ & $\num{1.22}$ & $\num{0.38}$ & $\num{0.247}$ & $\num{1.787}$ & $\num{17.50}$ & $\num{0.900}$ & $\num{3.700}$ & $\num{1.498}$ & $\num{4.331}$ & $\num{1.476}$ \\
& B & $\num{0.35}$ & $\num{0.56}$ & $\num{0.01}$ & $\num{0.332}$ & $\num{1.756}$ & $\num{27.18}$ & $\num{1.047}$ & $\num{5.764}$ & $\num{2.333}$ & $\num{6.744}$ & $\num{2.299}$\\
& C & $\num{0.50}$ & $\num{0.69}$ & $\num{0.01}$ & $\num{0.314}$ & $\num{2.589}$ & $\num{21.44}$ & $\num{0.979}$ & $\num{4.897}$ & $\num{1.983}$ & $\num{5.757}$ & $\num{1.962}$\\ \hline

T8 & A & $\num{0.39}$ & $\num{2.14}$ & $\num{0.39}$ & $\num{0.236}$ & $\num{2.037}$ & $\num{13.27}$ & $\num{0.874}$ & $\num{2.920}$ & $\num{1.404}$ & $\num{3.421}$ & $\num{1.357}$ \\
& B & $\num{0.35}$ & $\num{0.56}$ & $\num{0.01}$ & $\num{0.332}$ & $\num{1.756}$ & $\num{27.18}$ & $\num{1.047}$ & $\num{5.764}$ & $\num{2.771}$ & $\num{6.744}$ & $\num{2.675}$ \\
& C & $\num{0.55}$ & $\num{0.66}$ & $\num{0.02}$ & $\num{0.309}$ & $\num{2.646}$ & $\num{21.70}$ & $\num{0.983}$ & $\num{4.728}$ & $\num{2.273}$ & $\num{5.519}$ & $\num{2.189}$ \\ \hline

A6 & A & $\num{0.37}$ & $\num{1.21}$ & $\num{0.38}$ & $\num{0.243}$ & $\num{1.584}$ & $\num{20.97}$ & $\num{0.923}$ & $\num{3.638}$ & $\num{1.819}$ & $\num{4.281}$ & $\num{1.840}$ \\
& B & $\num{0.35}$ & $\num{0.50}$ & $\num{0.01}$ & $\num{0.332}$ & $\num{1.759}$ & $\num{27.35}$ & $\num{1.045}$ & $\num{5.849}$ & $\num{2.925}$ & $\num{6.943}$ & $\num{2.984}$\\
& C & $\num{0.45}$ & $\num{0.58}$ & $\num{0.01}$ & $\num{0.318}$ & $\num{2.380}$ & $\num{22.99}$ & $\num{0.987}$ & $\num{5.327}$ & $\num{2.664}$ & $\num{6.392}$ & $\num{2.748}$ \\ \hline

 \end{tabular}
\end{table*}

The classification of different solution types works as follows: for a fixed
value of $\epsilon_{\rm max}$, there exists a critical value of the degree of
differential rotational $\hat{A}^{-1}_{\rm crit}$, defined by the condition
\begin{equation}
  \label{eq:Acrit}
  \frac{\partial \hat{A}^{-1}}{\partial (r_p \slash r_e)} \biggr \rvert_{\epsilon_{\text{max}}} = 0 =
  \frac{\partial \hat{A}^{-1}}{\partial \hat{\beta}} \biggr \rvert_{\epsilon_{\text{max}}}\,.
\end{equation}
The function $\hat{\beta}(r_p\slash r_e)$ that corresponds to
$\hat{A}^{-1}_{\rm crit}$ for a given energy density is referred to as the
``separatrix''. This quantity partitions the solution space into four regions,
each corresponding to a different solution type. Models are classified directly
using the separatrix in the $(r_p \slash r_e, \hat{\beta})$ plane, as shown in
Figure~\ref{fig:Beta-rp}, where we show the separatrix for EOS A5 at energy
density $\SI{0.5e15}{\g\per\cm\cubed}$. Figure~\ref{fig:Beta-rp} is consistent
with the results observed for polytropic and hadronic EOSs \cite{Ansorg2009,
  Espino2019}. However, since we were not able to build type D stars, there are
three regions instead of four in Figure~\ref{fig:Beta-rp}.

In practice, instead of working directly with the separatrix, it is more
convenient to consider $\hat{A}^{-1}_{\rm crit}$ as a function of maximum energy
density in order to assign a type to a configuration. For a fixed EOS and
$\epsilon_{\rm max}$, all equilibrium models with
$\hat{A}^{-1} < \hat{A}^{-1}_{\rm crit}$ are of either type A or B, whereas
those with $\hat{A}^{-1} > \hat{A}^{-1}_{\rm crit}$ are of type C (or D). Then,
a solution is categorized as spheroidal if the maximum energy density occurs in
the center of mass. Since we do not have control over $\hat{\beta}$, we cannot
directly solve Equation~\eqref{eq:Acrit} to find the critical value for
$\hat{A}^{-1}$. Hence, as in \cite{Espino2019}, we estimate
$\hat{A}^{-1}_{\rm crit}$ with the maximum of the function
$\hat{A}^{-1}_{\rm min} (r_p/r_e)$, where $\hat{A}^{-1}_{\rm min}$ is the
smallest degree of differential rotation for which an equilibrium model exists
with the given maximum energy density. This method allows the calculation of
$\hat{A}^{-1}_{\rm crit}$ to within $\SI{1}{\percent}$ accuracy \cite{Espino2019}.

Figure~\ref{fig:separatrix} shows the critical degree of differential rotation
as a function of the maximum energy density for all the hybrid EOSs we consider
in this work. The figure exhibits features which are unique to hybrid EOSs but
which may be reconciled with the results of hadronic and strange-quark-matter
EOSs. Crucially, $\hat{A}^{-1}_{\rm crit}$ is not a monotonically decreasing
function of $\epsilon_{\rm max}$,\footnote{We refer to the increase and subsequent
  decrease of $\hat{A}^{-1}_{\rm crit}(\epsilon_{\rm max})$ at values of
  $\epsilon_{\rm max}$ above the phase transition (as shown in Figure
  \ref{fig:separatrix}) as a ``bump'' feature.} and, as a result, for some
values of $\hat{A}^{-1}$, an equilibrium sequence with varying
$\epsilon_{\rm max}$ (such as the ones considered in
Section~\ref{sec:universal-relations-1}) can jump back and forth between being
class A or B and C. The significant increases in mass that we find in
Section~\ref{sec:maxim-mass-class} correspond to a jump in models from
type A to type C.

In \cite{Ansorg2009}, homogeneous stars described by an incompressible EOS in which
\begin{equation}\label{eq:incompressible_eos}
\epsilon(P) = \epsilon_0\,,
\end{equation}
where $\epsilon_0$ is a constant, were considered. For such an EOS, the pressure and
energy density do not depend on one another, similar to the constant pressure
regions of the EOSs studied in this work. In \cite{Ansorg2009} it was found that
the function $\hat{A}^{-1}_{\rm crit} (\epsilon_{\rm max})$ for incompressible EOSs
first increases to a maximum and subsequently decreases as the energy density
increases. This same feature was observed for strange-quark-matter EOSs, which
may be suitably approximated as homogeneous bodies \cite{Szkudlared2019}. We
note that similar features appear in the curves presented in
Figure~\ref{fig:separatrix} above the values of energy density corresponding to
the constant pressure regions of the EOSs. It is possible that the ``bump''
features seen in the curves in Figure~\ref{fig:separatrix} arise for the same
reasons as those found in \cite{Ansorg2009} and \cite{Szkudlared2019} for
homogeneous and strange-quark-matter EOSs, respectively.

In Table~\ref{tab:maxmass_ABC} we list the maximum rest mass models found in our
search for each solution type we were able to construct with the Cook code and
for three representative EOSs. We focus on the T1, T8, and A6 EOSs to highlight
particular features of the solution space. As can be seen from
Figure~\ref{fig:separatrix}, the ``bump'' feature of the function
$\hat{A}^{-1}_{\rm crit} (\epsilon_{\rm max})$ results in the appearance of A
and B solutions at values of $\epsilon_{\rm max}$ and $\hat{A}^{-1}$ which are
high compared to hadronic EOSs. Despite the presence of type A and B stars in
larger regions of the parameter space compared to hadronic EOSs, we find that
the maximum rest mass configurations still correspond to modest degrees of
differential rotation and not for the largest energy densities considered. For
all three EOSs, we find that the type B models are the most massive, which
agrees with the findings from studies with hadronic EOSs \cite{Espino2019}. The
largest gain in mass is for EOS A6, with which it is possible to construct stars
almost {\it three times} more massive than the corresponding TOV limit. This
increase in rest mass is larger than the ones seen for the hadronic EOSs studied
in \cite{Espino2019}.  Note that the type B stars tend to have the lowest value
of $\epsilon_{\rm max}$ among the maximum rest mass solutions, suggesting that
these configurations predominantly sample from the low density part of the
corresponding EOS. Note also that for this reason the T1 and T8 type B models
presented in Table \ref{tab:maxmass_ABC} are the same. The T1 and T8 EOSs are
identical in the low density regime and begin to diverge at $ \epsilon \approx
\SI{0.66e15}{\g\per\cm\cubed}$, whereas the maximum rest mass stars for each of
these EOS have $ \epsilon_{\rm max} = \SI{0.56e15}{\g\per\cm\cubed}$. Because
the \emph{maximum} energy density of these configuration is the same, and,
because the EOSs are identical below this energy density, the corresponding
solutions are identical. Any EOS that is the same in the low density regime as
the T1 EOS will have an identical maximum rest mass type B model (including T2,
T4, T6, T7, T9, and TT). However, note that because of their different TOV
masses, each of these EOSs exhibit a different maximal increase in the rest mass
compared to the TOV limit rest-mass. The maximum rest mass type C solutions
presented in Table \ref{tab:maxmass_ABC} are different between T1 and T8,
because the corresponding values of $\epsilon_{\rm max}$ are greater than the
energy density at which the two EOS begin to diverge from one another. The
models presented in Table \ref{tab:maxmass_ABC} are not the definitively most
massive, but only the largest we were able to locate in the solution space. We
were unable to scan the entire solution space using the Cook code (in particular
we could not construct type D stars or highly pinched type B ones). Moreover,
because different solution types can emerge as the maximum energy density
changes, locating the absolute maximum rest mass models for the solution types
we were able to compute depends on the resolution in the parameters $
{\epsilon_{\rm max}, r_p/r_e, \hat{A}^{-1}}$ used to scan the solution space. We
used a step in $\epsilon_{\rm max}$ and $r_p/r_e$ of $\num{0.01}$ and a step in
$\hat{A}^{-1}$ of $\num{0.01}$ for the class A, but a step of $\num{0.05}$ for
families B and C.

\def\prd{Phys. Rev. D}\def\prl{Phys. Rev. Lett.}\def\apjl{Astrophys. J.
  Lett.}\def\apjs{Astrophys. J. Suppl.}\def\apj{Astrophys. J.}\def\aj{Astron.
  J.}\def\aap{Astron. Astrophys.}\def\aaps{Astron. Astrophys.
  Suppl.}\def\araa{Ann. Rev. of Astron. and Astrophys.}\def\adp{Ann.
  Phy.}\def\cqg{Classical Quant. Grav.}\def\mnras{Mon. Not. R. Astron.
  Soc.}\def\physrep{Phys. Rep.}\def\nat{Nat.}\def\pasj{Pub. Astron. Soc. of
  Jap.}\def\prc{Phys. Rev. C}\def\sovast{Soviet. Ast.}

\providecommand{\href}[2]{#2}\begingroup\raggedright\endgroup

\end{document}